\documentclass[trackchanges]{aastex701}

  \renewcommand{\d}{\mathrm{d}}

  \newcommand{\vct}[1]{\boldsymbol{#1}}
  \newcommand{\wt} [1]{\widetilde{#1}}
 \newcommand{\z}{\boldsymbol{z}}    
  \newcommand{\Rey}{\mathrm{Re}}

\usepackage{amsmath}
\usepackage{amssymb}
\usepackage{bm}
\usepackage{siunitx}
\usepackage{float}

\begin{document}

\title{Angular dependence of third-order law in anisotropic MHD turbulence\footnote{Footnotes can be added to titles}}

\author[0000-0002-4858-0505,gname=Bin,sname=Jiang]{Bin Jiang}
\affiliation{School of Mechanical Engineering and Mechanics, Xiangtan University, Xiangtan 411100, PR China}
\affiliation{Guangdong Provincial Key Laboratory of Turbulence Research and Applications, Department of Mechanics and Aerospace Engineering, Southern University of Science and Technology, Shenzhen 518055, PR China}
\affiliation{Guangdong-Hong Kong-Macao Joint Laboratory for Data-Driven Fluid Mechanics and Engineering Applications, Southern University of Science and Technology, Shenzhen 518055, PR China}
\email{jiangbin@xtu.edu.cn}

\author[gname=Zhuoran,sname=Gao]{Zhuoran Gao} 
\affiliation{Department of Physics and Astronomy, University of Delaware, DE 19716, USA}
\email{veciam@udel.edu}

\author[orcid=0000-0003-2965-7906,gname=Yan,sname=Yang]{Yan Yang} 
\affiliation{Department of Physics and Astronomy, University of Delaware, DE 19716, USA}
\email[show]{yanyang@udel.edu}
\correspondingauthor{Yan Yang}

\author[orcid=0000-0003-4168-590X]{Francesco Pecora} 
\affiliation{Department of Physics and Astronomy, University of Delaware, DE 19716, USA}
\email{fpecora@udel.edu}

\author[]{Kai Gao}
\affiliation{Guangdong Provincial Key Laboratory of Turbulence Research and Applications, Department of Mechanics and Aerospace Engineering, Southern University of Science and Technology, Shenzhen 518055, PR China}
\affiliation{Guangdong-Hong Kong-Macao Joint Laboratory for Data-Driven Fluid Mechanics and Engineering Applications, Southern University of Science and Technology, Shenzhen 518055, PR China}
\email{12131114@mail.sustech.edu.cn}

\author[]{Cheng Li}
\affiliation{Guangdong Provincial Key Laboratory of Turbulence Research and Applications, Department of Mechanics and Aerospace Engineering, Southern University of Science and Technology, Shenzhen 518055, PR China}
\affiliation{Guangdong-Hong Kong-Macao Joint Laboratory for Data-Driven Fluid Mechanics and Engineering Applications, Southern University of Science and Technology, Shenzhen 518055, PR China}
\email{licheng1126@gmail.com}

\author[0000-0002-2814-7288,gname=Sean,sname=Oughton]{Sean Oughton}
\affiliation{Department of Mathematics, University of Waikato, Hamilton 3240, New Zealand}
\email{sean.oughton@waikato.ac.nz}

\author[0000-0001-7224-6024,gname=William,sname=Matthaeus]{William H. Matthaeus}
\affiliation{Department of Physics and Astronomy, University of Delaware, DE 19716, USA}
\email{whm@bartol.udel.edu}

\author[0000-0001-5891-9579,gname=Minping,sname=Wan]{Minping Wan}
\affiliation{Guangdong Provincial Key Laboratory of Turbulence Research and Applications, Department of Mechanics and Aerospace Engineering, Southern University of Science and Technology, Shenzhen 518055, PR China}
\affiliation{Guangdong-Hong Kong-Macao Joint Laboratory for Data-Driven Fluid Mechanics and Engineering Applications, Southern University of Science and Technology, Shenzhen 518055, PR China}
\email[show]{wanmp@sustech.edu.cn}
\correspondingauthor{Minping Wan}

\collaboration{all}{}

\begin{abstract}
In solar wind turbulence, the energy transfer/dissipation rate is typically estimated using MHD third-order structure functions calculated using spacecraft observations. 
However, the inherent anisotropy of solar wind turbulence leads to significant variations in structure functions along different observational directions, thereby affecting the accuracy of energy-dissipation rate estimation. 
An unresolved issue is how to optimise the selection of observation angles under limited directional sampling to improve estimation precision. 
We conduct a series of MHD turbulence simulations with different mean magnetic field strengths, $ B_0 $. 
Our analysis of the third-order structure functions reveals that the global energy dissipation rate estimated around a polar angle of $ \theta = 60^\circ$ agrees reasonably with the exact one for $ 0 \le B_0/b_{rms} \le 5 $, where $b_{rms}$ denotes the root-mean-square magnetic field fluctuation. 
The speciality of $60^\circ$ polar angle can be understood by the Mean Value Theorem of Integrals, since the spherical integral of the polar-angle component ($\wt{T_\theta}$) of the divergence of Yaglom flux is zero, and $\wt{T_\theta}$ changes sign around 60$^\circ$. Existing theory on the energy flux vector as a function of the polar angle is assessed, and supports the speciality of $60^\circ$ polar angle. 
The angular dependence of the third-order structure functions is further assessed with virtual spacecraft data analysis. 
The present results can be applied to measure the turbulent dissipation rates of energy in the solar wind, which are of potential importance to other areas in which turbulence takes place, such as laboratory plasmas and astrophysics.

\end{abstract}



\section{Introduction}
\label{sec:intro}

 Magneto-hydrodynamic (MHD) turbulence, characterised by high Reynolds numbers, is  ubiquitous in natural environments, including the solar wind 
 \citep{coleman1968turbulence,jokipii1970interplanetary, Parker-cmf,  matthaeus1982measurement, tu1995mhd, BrunoCarboneLRSP13}. The solar wind, a steady flow of charged particles that fills the heliosphere, plays a significant role in space weather and satellite operation. Therefore, the study of solar wind is vital for both scientific inquiry and engineering solutions. 
At scales larger than the kinetic ones, MHD turbulence is capable of modeling solar wind fluctuations. 
An important parameter in the study of turbulence is the energy dissipation rate 
 ($\varepsilon$), 
which also measures the intensity of the cascade process. Understanding the energy cascade in MHD turbulence is crucial to understand the heating phenomena in solar wind \citep{matthaeus1982measurement}.
However, it is not currently possible to directly measure 
   $\varepsilon$ (or its cousins) 
in the solar wind. In space environments, plasmas generally act as (almost) collisionless systems, so that 
viscosity and resistivity may be difficult to even define,
and thus direct measurements of viscous and resistive dissipation rates are typically not feasible
  \cite[cf.][]{YangEA24-effvisc, AdhikariEA25-effvisc, AdhikariEA25-SWvisc}.

An alternative approach is to estimate the energy dissipation rates \emph{in}directly, based on the MHD third-order law, 
obtained by extending Kolmogorov theory to MHD turbulence \citep{politano1998karman,PolitanoPouquet98-grl}.
Assuming isotropic MHD turbulence
\citep{politano1998karman,Sorriso2007,macbride2008turbulent,stawarz2009turbulent},
one obtains the \emph{isotropic third-order law}:
\begin{equation}
   Y_\ell^\pm(\ell) 
      \stackrel{\text{def}}{=}
    \langle \delta z_\parallel^{\mp} (\delta\vct{z}^{\pm})^2 \rangle 
    = 
     - \frac{4}{3} \bar{\varepsilon}^{\pm} \ell.
\label{eq:SH_MHD} 
\end{equation}
In a standard notation, $ \vct{z}^{\pm} = \vct{u}\pm \vct{b} $ 
are the Els\"asser variables 
with $\vct{u}$ the velocity fluctuation field
and $\vct{b}$ the fluctuating magnetic field in Alfv\'en units 
 (i.e., $\vct{b}$ is scaled by $\sqrt{4\pi\rho}$ 
  with $\rho$ the uniform mass density). 
The Els\"asser increments are defined as $\delta\vct{z}^{\pm} (\vct{x},\vct{\ell})= \vct{z}^{\pm} (\vct{x}+\vct{\ell}) - \vct{z}^{\pm} (\vct{x})$ and $\delta z_\parallel^{\pm}=\delta\vct{z}^{\pm} \cdot\vct{\ell}/\ell $,
for general position vector $ \vct{x} $ and lag $ \vct{\ell} $. $\vct{\ell}$ vector in the case of solar wind measurements is the 
``longitudinal'' direction aligned with the mean flow.
Here, $\bar{\varepsilon}^{\pm}$ represent the globally averaged dissipation rates of Els\"asser energies,
  $ \langle | \vct{z}^\pm(\vct{x}) |^2 \rangle /2 $.

An important limitation of 
 Eq.~\eqref{eq:SH_MHD}
is that it relies on the isotropic assumption, which is often violated in MHD systems with a mean magnetic field, such as the solar wind.
Turbulent systems threaded by a mean magnetic field, $\vct{B}_0$     
inherently display anisotropy \citep[e.g.,][]{shebalin1983anisotropy,Oughton_1994,matthaeus1996anisotropic,horbury2008anisotropic,luo2010observations,Oughton2020CB}.
This anisotropy introduces complications regarding measurement of energy transfer, that future solar wind missions such as HelioSwarm will seek to address \citep{SpenceEA19,MatthaeusEA19-whitepaper,retino2022particle_PO,marcucci2024PO}.

Anisotropic aspects of third-order laws have been investigated in several MHD simulation studies.
\cite{Verdini_2015} demonstrated that the isotropic third-order law, 
  when evaluated at a fixed polar angle $\theta$, 
underestimates the globally averaged dissipation rates for $\theta$ below $40^\circ$ 
  and overestimates them beyond $70^\circ$. 
Thus, for the purposes of estimating dissipation rates,
there may be an optimal polar angle between $40^\circ$ and $70^\circ$. 
In an investigation of the anisotropic energy transfer induced by the mean magnetic field,
 \cite{jiang2023energy} 
found that the isotropic third-order law evaluated at  
 $\theta = 60^\circ$ 
 matches the fully direction-averaged third-order law 
   (see Section~\ref{sec:theory} for definitions) 
to better than 5\%. 
Whereas, at other values of $\theta$ the discrepancy can be as high as 45\%.
 An aim of the present work is to
confirm and understand this speciality of the $60^\circ$ polar angle.

Unfortunately, in practice \emph{in situ} spacecraft observations typically
lack full directional coverage,
limiting the directional averaging that may be performed
\citep{Osman2011Anisotropic}.
Consequently, many of the studies using solar wind data to calculate longitudinal third-order moments have in essence employed Eq.~\eqref{eq:SH_MHD} at a fixed polar angle.
Single-point measurements with clear evidence of linear scaling for the third-order moment were first presented by \cite{Sorriso2007}. They also found that there are some regions without such isotropic third-order law scaling, which they suggested was mainly due to the anisotropy associated with the high-latitude regions, with compressibility and inhomogeneity as secondary factors. 
Based on the isotropic third-order law, \cite{macbride2008turbulent} adopted a 1D + 2D method to estimate the dissipation rate, reducing the estimation error. However, \cite{stawarz2009turbulent} argued that statistical convergence can significantly affect third-order scaling, and claimed that it requires about a year of ACE data at 64\,s cadence to reduce the fractional error of the dissipation rate estimation below $30\%$.  
The majority of previous works have leaned towards the use of the isotropic third-order law, with several exceptions. For example, \cite{Osman2011Anisotropic} and \cite{Bandyopadhyay_2018} used \emph{Cluster} \citep{escoubet2001cluster} and \emph{MMS} (Magnetospheric Multiscale mission) \citep{burch2016MMS}
data, respectively, with the directional averaging technique to estimate the dissipation rates. More details regarding such observational studies can be found in the review paper \cite{Marino2023}. 
\cite{Percora2023PRL} employed \emph{MMS} data and the LPDE (Lag Polyhedral Derivative Ensemble) method \cite{Pecora2023ApJ} to estimate the divergence of the Yaglom flux in lag space directly, and attain more accurate and robust estimations for the dissipation rate, without making any assumption about isotropy.  
It is clear that studying the angular dependence of the third-order law is meaningful, especially in attempting to find an optimized angle for employing spacecraft measurements. 

The layout of the paper is as follows.
Section \ref{sec:theory} provides a concise overview of the third-order law.
Section \ref{sec:numerical_configs} presents the numerical method, covering the simulation configurations, and some important characteristics of the (statistically steady) flow and magnetic fields.  
In Section \ref{sec:results}, the examination of the energy flux, including its vector properties and divergence, will be studied to explore the speciality of the $60^\circ$ polar angle. Existing theoretical relationships between the polar-angle and energy flux components will be verified. Further, the effects of the azimuthal average will be assessed. Finally, the angular dependence will be explored with virtual spacecraft data analysis. The conclusions are summarised at the end.

    \section{Third-order law}\label{sec:theory}

This section outlines three different averaging methods pertinent to the third-order law, as discussed in \cite{jiang2023energy}. Beginning with the von K\'arm\'an--Howarth  (vKH) equation 
 \citep{Karman1938statistical,MoninYaglom,FrischBook95,PolitanoPouquet98-grl}, 
we employ a general expression for an MHD third-order law. 
In an inertial range 
 (i.e., the scales for which the non-stationary and dissipative terms in the vKH equation are negligible), 
the cross-scale energy transfer may be expressed as
\begin{equation}
  \nabla_l \cdot \vct{Y}^{\pm} 
 = 
  \nabla_l \cdot \langle \delta\vct{z}^{\mp} |\delta\vct{z}^{\pm}|^2 \rangle 
 = 
   -4 \varepsilon^{\pm}, 
 \label{eq:div_3rd_order_law}
\end{equation}
where 
 $ \vct{Y}^{\pm} (\vct{\ell} )
    = 
    \langle 
       \delta\vct{z}^{\mp} |\delta\vct{z}^{\pm}|^2 
    \rangle $ 
are third-order structure functions, also called the Els\"asser
  energy or Yaglom flux vectors,
 and angle brackets 
   $ \langle \cdot \rangle$ 
denote ensemble averaging 
 (assumed equal to averaging over $\vct{x}$ for simulation data). 
As noted earlier, the Els\"asser variables are
  $\vct{z}^{\pm} = \vct{u} \pm \vct{b}$, 
with increments
  $ \delta\vct{z}^{\pm} (\vct{x},\vct{l})= \vct{z}^{\pm} (\vct{x}+\vct{l}) - \vct{z}^{\pm} (\vct{x})$.
Our simulations will employ hyper-dissipation so that explicit forms for the mean dissipation rates of Els\"asser energies are $\varepsilon^{\pm}=\nu_h  \sum\limits_{\vct{k}} k^{2h} \langle |\hat{\vct{z}}^{\pm} (\vct{k},t)|^2 \rangle $, 
where $h$ denotes the hyper-viscosity index and $\nu_h$ denotes the hyper-viscosity coefficient, taken equal to the hyper-resistivity.
The gradient $\nabla_\ell$ operates in lag space (i.e., wrt the coordinates of the lag vector $\vct{\ell}$).  In particular, in spherical coordinates we have:
\begin{eqnarray}
 &&\nabla_\ell \cdot \vct{Y}^{\pm}
 = 
 \frac{1}{\ell^2} \frac{\partial (\ell^2 {Y_\ell^\pm})}{\partial \ell}  
 + 
 \frac{1}{\ell \sin\theta}  \frac{\partial(\sin\theta \, {Y_\theta^\pm})}{\partial \theta} 
 + 
 \frac{1}{\ell \sin\theta}  \frac{\partial ({Y_\phi^\pm})}{\partial \phi} 
 = {T_\ell^\pm} + {T_\theta^\pm} + {T_\phi^\pm}
 ,
\label{eq:Yvector_components}   
\end{eqnarray}
where, for convenience of description, the rightmost terms, $T_\ell^\pm$, $T_\theta^\pm$, and $T_\phi^\pm$, are a shorthand for the contributions to this divergence. 
We take the $z$ axis to be in the direction of the external mean field, $ {\vct{e}}_z$,
so that $\theta$ is the angle between the lag vector and $\vct{B}_0$.
Note that because the turbulence is homogeneous and Eq.~\eqref{eq:div_3rd_order_law} involves (ensemble) averaging,  
there is no dependence of $\vct{Y}^{\pm}$ on position $\vct{x}$, and the $\varepsilon^{\pm}$ are independent of both position $\vct{x}$ and lag $\vct{\ell}$. 

Eq.~\eqref{eq:div_3rd_order_law} 
can be reformulated 
in terms of various levels of angle averaging. 
First, taking a volume integral over a sphere with radius $\ell = | \vct{\ell} |$ yields
\begin{equation}
  \iiint_{|\vct{\ell}|\le \ell} \nabla_\ell \cdot \vct{Y}^{\pm} \, \d{V}
  = 
  \iiint_{|\vct{\ell}|\le \ell} -4 \varepsilon^{\pm} \, \d{V}
  =
  -\frac{16 \pi}{3} \varepsilon^{\pm} \ell^3
   .
\label{eq:div02_3rd_order_law}
\end{equation}
Using Gauss's theorem, this can be written as a surface integral,
\begin{equation}
   \oint_{|\vct{\ell}|=\ell} Y_\ell^{\pm} \, \d{S} 
 = 
   -\frac{16 \pi}{3} \varepsilon^{\pm} \ell^3, 
\label{eq:div03_3rd_order_law}
\end{equation}
where the longitudinal third-order structure functions
    $ Y_\ell^{\pm} =\langle \delta z_\ell^{\mp} |\delta\vct{z}^{\pm}|^2 \rangle$
is the projection of the energy flux vectors along
  $\vct{\ell}$
  and 
  $\delta z_\ell^{\mp} = \delta\vct{z}^{\mp} \cdot \frac{\vct{\ell}}{\ell} $.
Using spherical coordinates we may express this 
in terms of the solid angle average of $Y^\pm_\ell$:
\begin{equation}
  \frac{1}{4\pi} \int_{0}^{2\pi} \int_{0}^{\pi} 
         Y_\ell^{\pm}  \sin\theta \, \d\theta \d\phi 
   = 
   - \frac{4}{3} \varepsilon^{\pm} \ell, 
 \label{eq:3d_3rd_order_law}
\end{equation}
where $\theta$ represents the polar angle (from the $\vct{B}_0$ axis) and $\phi$ the azimuthal angle. 
Considering that no assumptions about rotational symmetry are made in going from 
Eq.~\eqref{eq:div_3rd_order_law} to 
Eq.~\eqref{eq:3d_3rd_order_law}, the physical content of 
Eq.~\eqref{eq:3d_3rd_order_law} is as general as the derivative form Eq.~\eqref{eq:div_3rd_order_law}. 
{The full generality of 
Eq.~\eqref{eq:3d_3rd_order_law} follows from the rigorous theorem given by \cite{NieTanveer99} and restated in more accessible terms by \cite{TaylorEA03} and \cite{wang2022strategies}.} However, Eq.~\eqref{eq:3d_3rd_order_law} is simpler in the sense that accurate determination of integration only requires the longitudinal component of the energy flux vectors, $Y_\ell^{\pm}$, 
on the spherical surface spanned by the coordinates $(\theta, \phi)$ in the 3D lag space.   

The most general form of $Y_\ell^{\pm}$ should be a function of $\ell$, $\theta$ and $\phi$, that is, $Y_\ell^{\pm}(\ell,\theta,\phi)$. 
The theory of tensor invariants may be used to obtain general constraints on the functional form for $Y^\pm_\ell$
  \citep[e.g.,][]{Robertson40,BatchelorTHT,podesta2007anisotropic}.
Previous studies have typically restricted themselves to the purely isotropic assumption ($ Y_\ell^{\pm}$ independent of $\theta$ and $\phi$) or treated anisotropic turbulence with azimuthal symmetry ($Y_\ell^{\pm}$  independent of $\phi$) as implemented, for example, by \cite{stawarz2009turbulent}.  
Assuming isotropy, Eq.~\eqref{eq:3d_3rd_order_law}  reduces to:
\begin{equation}
\frac{1}{4\pi} \; Y_{\ell,{\rm iso}}^{\pm}(\ell) 
       \int_{0}^{2\pi} \int_{0}^{\pi}  \sin\theta \, \d\theta \d\phi 
 \;\; = \;\; 
   Y_{\ell,{\rm iso}}^{\pm}( \ell) 
= 
   -\frac{4}{3} \varepsilon^{\pm} \ell , 
\label{eq:iso_3rd_order_law}
\end{equation}
which we refer to as the \emph{isotropic (third-order) law}.
This is the same as Eq.~\eqref{eq:SH_MHD}.

To better understand the anisotropic energy transfer in the inertial range, we provide here a systematic study of the dependence of the (longitudinal) structure function 
  $ Y_\ell^{\pm} (\ell, \theta, \phi)   $ 
on $\theta$ and $\phi$ for situations with different guide field magnitudes. 
Since simulation (and spacecraft) data are only available at discretely spaced points, we will form estimates of $Y^\pm_\ell$ using discrete sets of angles.  
Specifically, to cover the sphere we elect to use lag vectors in 37 directions, uniformly spaced in azimuthal and polar angles ($\Delta\theta=15^\circ$ and $\Delta\phi=60^\circ$).
Note that for $\theta=0^\circ$ the azimuthal angle does not play a role, since it is formally undefined. 
Moreover, due to the uniform external mean magnetic field, we may restrict the range of $\theta$ to $[0^\circ, 90^\circ]$.
Thus we have $ N_i = 7$ polar angles and $ N_j = 6$ azimuthal angles
 (except for $ \theta = 0$ when $ N_j = 1 $). 
To ensure accuracy when calculating the divergence of the energy flux in Section~\ref{sec:Eflux} and the fitting of structure function with varying $\theta$ in model verification in Section~\ref{sec:Similarity}, we employ a finer grid with  
$ N_i = 23$ uniformly spaced polar angles. 
Also, as the divergence involves factors of $\sin\theta$ in several denominators, we drop the $\theta = 0 $ points from these calculations.
A 3D Lagrangian interpolation is used to estimate data values located between grid points.

In order of increasing levels of angle-averaging, the three estimates for $Y_\ell^{\pm}$ we consider are:

I) $Y^\pm_\ell$ is evaluated at discrete pairs of angles, $\theta_i$ and $\phi_j$, but with no averaging over angles: 
\begin{equation}
   Y_\ell^{\pm}( \ell, \theta_i,\phi_j) .
\label{eq:3rd_order_sf_3D_lag} 
\end{equation} 

This represents a local radial (or longitudinal) energy transfer, where `local' means at the specific azimuthal and polar angles. 
The total radial energy transfer at scale $\ell$ is the sum of the Eq.~\eqref{eq:3rd_order_sf_3D_lag} contributions from all azimuthal $\phi_j$ and polar $\theta_i$ directions at the same lag length.  
Separate estimates are made for each of the 37 directions, that is, 
  $ Y_\ell^{\pm}( \ell, \theta_i,\phi_j)$ 
is calculated for $\theta_i\in[0^\circ:15^\circ:90^\circ]$ 
and $\phi_j\in[0^\circ:60^\circ:300^\circ]$; 
the middle value in the square brackets indicates the step size for the angle.

II) The azimuthally averaged form of the third-order structure function,
\begin{equation}
\wt{ Y_\ell^{\pm}}{(\ell, \theta_i)} 
 = 
 \frac{1}{2\pi}\int_{0}^{2\pi} Y_\ell^{\pm}(\ell, {\theta_i},\phi) \, \d\phi
 \approx 
\frac{\sum_{j=1}^{N_j} Y_\ell^{\pm}(\ell, \theta_i,\phi_j) }{N_j},
\label{eq:3rd_order_sf_2D_lag}
\end{equation} 
describes the (polar) anisotropy of local radial energy transfer, where here `local' means at a specific polar angle. The total transfer rate is the sum over all $ \theta_i $ 
of the contributions given by Eq.~\eqref{eq:3rd_order_sf_2D_lag} at a given $\ell$.
In practice, this estimate just makes appropriate averages over $\phi_j$ of the estimates determined using Method I, Eq.~\eqref{eq:3rd_order_sf_3D_lag}. 

III) The 
  `full'\footnote{Here `full' means with respect to the discrete $\theta_i$, $\phi_j$ grid, rather than over the continuum of angles.} 
direction-averaged form of the third-order structure function,
\begin{align}
 \overline{ Y_\ell^{\pm}}(\ell)
  =
  \frac{1}{4\pi}\int_{0}^{2\pi} \int_{0}^{\pi} Y_\ell^{\pm} \sin\theta \, \d\theta \d\phi 
 \approx  
 \frac{\sum_{j=1}^{N_j} \sum_{i=1}^{N_i}  
   Y_\ell^{\pm} (\ell, \theta_i,\phi_j) 
    \sin\theta_i}{N_j\sum_{i=1}^{N_i} \sin\theta_i}
 .
 \label{eq:3rd_order_sf_1D_lag}
\end{align}
Recall, $N_i$ (=7) and $N_j$ (=6, usually) indicate the number of $\theta_i$ and $\phi_j$ angles used, respectively. This direction-averaged version of 
the (longitudinal) third-order structure function 
is formulated straight from the vKH equation---without assuming statistical isotropy---and is solely dependent on the lag length $\ell$. 
Assuming that the conditions needed for   Eq.~\eqref{eq:3d_3rd_order_law} 
  (and Eq.~\eqref{eq:div_3rd_order_law})
to hold apply, normalising $\overline{Y_\ell^{\pm}}$ with the factor $-4\ell/3$ provides an estimate of the cross-scale energy transfer rate (or energy dissipation rate), $\varepsilon$. 
Additionally, the range of scales over which this linear scaling with $\ell$ holds may be used to estimate the range of scales comprising the inertial range.

These three variants of the third-order structure functions enable approximations of the true energy dissipation rates, $\varepsilon^\pm$. 
For example, Method I, Eq.~\eqref{eq:3rd_order_sf_3D_lag}, is commonly employed in observational studies involving a single spacecraft. 
Our focus herein is on the \emph{total} energy dissipation rate, so that for each of the three forms 
  $ Y_\ell^{\pm}$, 
  $\wt{ Y_\ell^{\pm}}$, and 
  $\overline{ Y_\ell^{\pm}}$, 
we average their $+$ and $-$ components. 
These are then normalised by $ -\frac{4}{3} \varepsilon \ell $, where 
 $ \varepsilon = (\varepsilon^+ + \varepsilon^-)/2 
               =  \varepsilon_b + \varepsilon_v $
signifies the total energy cascade rate.
This yields, for example,
    $ -3(\wt{Y^+_\ell} + \wt{Y^-_\ell}) / (8 \varepsilon \ell) $,
which should be unity for inertial range scales.
As our simulations have low cross helicity, we expect statistical equality of the $+$ and $-$ components so that averaging them should not make much difference to the $\varepsilon$ estimates.  The situation in the inner heliosphere solar wind is different, with significant cross helicity levels present there.

\section{\label{sec:config}Numerical configurations} \label{sec:numerical_configs}

We solve the 3D incompressible MHD equations numerically, in Fourier space,
using a pseudo-spectral method, incorporating the two-thirds rule for dealiasing \citep{Orszag1971,Orszag1972,gottlieb1977numerical,Orszag_Tang_1979,canuto2007spectral}. Our simulations are conducted in a cubic domain $[0,2\pi)^3$, with periodic boundary conditions and the second-order Adams--Bashforth method for time integration. 
An external force $\vct{f}_v$ is included in the momentum equation\citep{yang2021effects}.  
This forcing is exerted solely on the first two wave number shells ($k=1, 2$), to support development of a wide inertial range; here $k=|\vct{k}|$. 

\begin{figure}[H]
	\centering
\includegraphics[width=0.8\linewidth,keepaspectratio]{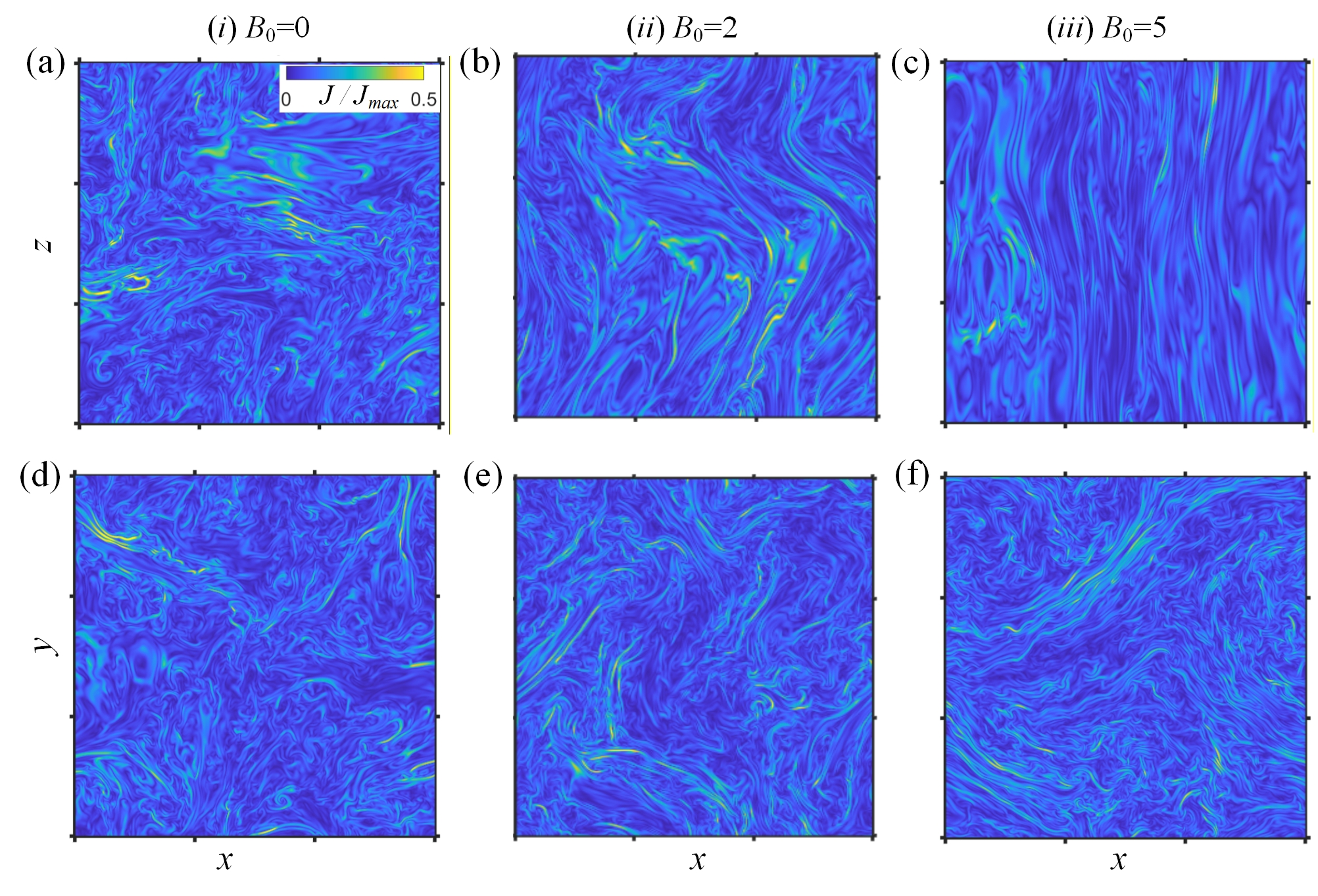}	\\[0.2cm]
 \caption{Current intensity ($J=|\nabla\times \textbf{b}|$) in domain cross-sections that are 
 (a,b,c) parallel--perpendicular ($z$--$x$) and (d,e,f) strictly perpendicular ($x$--$y$) with respect to the mean magnetic field, 
 for $B_0=$ 0 (panels a,d), 2 (b,e), and 5 (c,f).
 Note the increasing alignment of structures with the mean field direction seen in the top row.
 }
\label{fig:structures}
\end{figure}

Some key observables for the simulations are listed in Table~\ref{table:setup}.
These variables have been averaged over time; see \cite{jiang2023energy}
for further details. 
The anisotropic effects on the coherent structures can be illustrated directly using the current intensity magnitude, $J=|\nabla\times\vct{b}|$; see Figure~\ref{fig:structures}. With increasing $B_0$, the structures tend to be more elongated along the mean magnetic field direction 
 ($\vct{e}_z$), 
while the dynamics in the perpendicular plane remain statistically isotropic. Furthermore, the turbulent length-scale for the parallel ($\vct{B}_0$) direction
is obviously larger than that for the perpendicular direction, consistent with 
established results
  \citep[e.g.,][]{RobinsonRusbridge71,ZwebenEA79,zikanov1998direct,WeygandEA09,RuizEA11}.

We note that the values of the cascade rates, $\varepsilon_{v}$ and $ \varepsilon_{b}$, are strongly influenced by the energy injection from the forcing term.
For the present study, we focus on a steady-state analysis, for which the average energy transfer/cascade rates are equal to the dissipation rates, allowing us to use the same symbol(s), $\varepsilon^{\pm}$, for these conceptually distinct quantities.
\begin{table*}
   \centering   
   \caption{\label{table:setup}Numerical configuration parameters and steady-state values for some key quantities. 
   The external mean magnetic field strength is $B_0$. 
   $h$ denotes the hyper-viscosity index, and
   $\varepsilon_{v}$ and $\varepsilon_{b}$ represent the kinetic and magnetic dissipation rates, respectively. 
   $\Rey_{\lambda,v}$ and $\Rey_{\lambda,b}$ are the kinetic and magnetic Taylor Reynolds numbers;
   $k_\text{max} \eta_{k,v}$ illustrates the grid resolution, where $\eta_{k,v}$ and $k_\text{max}$ are respectively the Kolmogorov lengthscale in velocity field and the maximum resolved wavenumber (a third of the total grid in one direction); $T_e$ refers to the 
   large-eddy turnover time during the statistically steady period for the $B_0=0$ case. {$b_\text{rms}$ denotes the r.m.s. magnetic field fluctuation in the stationary period. $\theta_v$ describes the anisotropy angle for the velocity field as introduced in \cite{shebalin1983anisotropy}. 
   }} 
	\begin{tabular}{lccccccccccc} \hline
		$B_0$ & $h$ & Grids  &  $\varepsilon_v$ & $\varepsilon_b$ & {$b_\text{rms}$} & $\Rey_{\lambda,v}$ & $\Rey_{\lambda,b}$ & $k_\text{max} \eta_{k,v}$ &$\theta_v(^\circ)$ & Averaging period ($T_e$)        \\ \hline 
        0 & 2 &	512$^3$ & 0.67 & 1.24 & {0.92} & 846 & 255 & 1.73 & 55& [15:30]\\   
        2 & 2 &	512$^3$ & 0.77 & 1.12 & {1.14} & 951 & 473 & 1.71 & 72& [8:68]\\ 
        5 & 2 &	512$^3$ & 0.83 & 1.02 & {1.08} & 2415 & 435 & 1.70 & 83& [200:275]  \\ \hline     
\end{tabular}
\end{table*}

    \section{Results}
    \label{sec:results}

In this section, the angular dependence of the third-order law will be assessed, using the three methods described in Section~\ref{sec:theory}, and the speciality of the 60$^\circ$ polar angle will be highlighted. 
An analysis of the divergence of the energy flux vector provides a partial explanation for why $\theta \approx 60^\circ$ has elevated relevance. 
This will be followed by a discussion of how hints from this analysis provide insight into the modeling of the energy flux vector in axisymmetric MHD turbulence with external mean magnetic fields. 
Associated with this, the dependence of the third-order law on azimuthal angle will also be explored. Finally, the above findings regarding angular dependence will be verified using virtual spacecraft measurements.

    \subsection{Third-order law along 
     \texorpdfstring{$ 60^\circ $}{60 deg} 
     polar angle}\label{sec:observation}

\begin{figure}[H]
	\centering
	\includegraphics[width=0.8\linewidth,keepaspectratio]{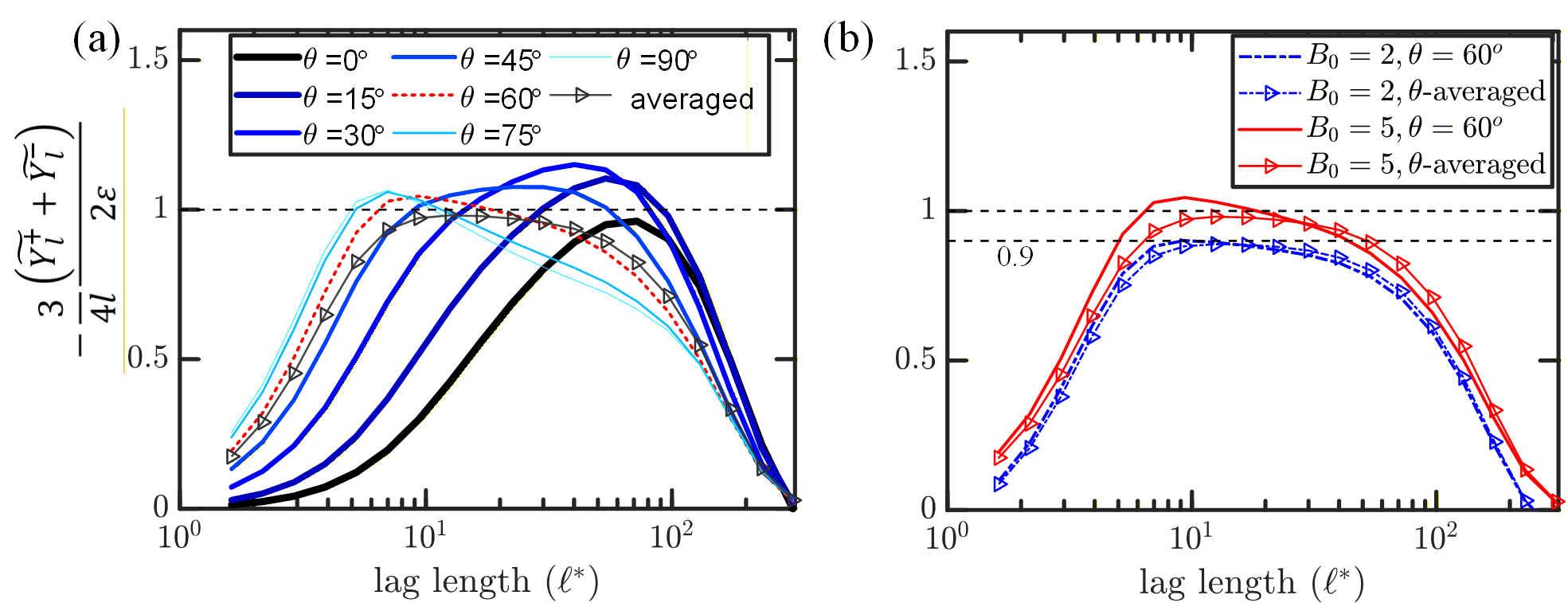}	
	\caption{Normalized longitudinal third-order structure functions, $ Y_\ell=(Y^+_\ell + Y^-_\ell)/2$, for various types of angle averaging.  
    (a) Azimuthal average using method II, $\wt{ Y_\ell}$ in         Eq.~\eqref{eq:3rd_order_sf_2D_lag}, for the $B_0=5$ simulation at indicated values of $\theta$. The solid line with triangles represents the average of these and is equivalent to the direction-averaged profile obtained via method III, Eq.~\eqref{eq:3rd_order_sf_1D_lag}:  
     $-3(\overline{ Y_\ell^+}+\overline{ Y_\ell^-})/({8\varepsilon \ell})$. (b) Direction-averaged (Eq.~\eqref{eq:3rd_order_sf_1D_lag}, lines with triangle symbols) 
    and azimuthally averaged at fixed $\theta = 60^\circ$ (Eq.~\eqref{eq:3rd_order_sf_2D_lag}, lines without triangle symbols) 
    for the $B_0=2$ and 5 simulations.
    Recall $\theta$ is the angle between the lag vector and the external mean magnetic field, $ B_0 \vct{e}_z $. 
    Time-averaging has been employed in all cases; see Table~\ref{table:setup}. In panel (b), the results for $B_0=2$ are shifted by -0.1 to avoid visual clutter with the $B_0=5$ results.} 
	\label{fig:averaged_fixed_theta}
\end{figure}

We begin by demonstrating the speciality of the third-order law along 
    $ \theta = 60^\circ $ empirically. 
Figure~\ref{fig:averaged_fixed_theta} displays normalised longitudinal third-order moments after azimuthal averaging, 
  $ \wt{Y_\ell}(\ell, \theta_i) $, and also after `full' directional averaging,
    $ \overline{Y_\ell}(\ell) $.
From Figure~2a, we see that the estimated energy dissipation rate peaks at large scales for parallel angles and at progressively smaller scales for more perpendicular angles. We also see that the maximum dissipation rate in the perpendicular plane  
  ($ \theta = 90^\circ $)
is larger than that in the parallel plane
   ($ \theta = 0^\circ $), 
consistent with other results 
  \citep[e.g.,][]{macbride2008turbulent,Verdini_2015}.
Figure~2b reveals the intriguing result that the 
    (fully) 
direction-averaged profiles are well approximated by the azimuthally averaged profile for a specific polar angle, namely $\theta = 60^\circ $, 
i.e.,
    $ \wt{Y_\ell}(\theta\approx 60^\circ) \approx \overline{Y_\ell} $.
However, the quality of this approximation decreases with increasing $B_0$. Specifically, the maximum deviation on the plateau is approximately 1\% when $B_0=2$ and increases to approximately 8\% when $B_0=5$. 
Standard Laplacian dissipation and hyper-dissipation simulations behave similarly in this regard
 (see Figure~\ref{fig:Yl_divY_Norm_B5_V5} of the Appendix). Detailed explanation about the $\theta = 60^\circ $ polar angle will be provided in Sections~\ref{sec:Eflux} and \ref{sec:Similarity}. 

    \subsection{Divergence of energy flux vector}
    \label{sec:Eflux}

\begin{figure}
	\centering
	\includegraphics[width=0.85\linewidth,keepaspectratio]{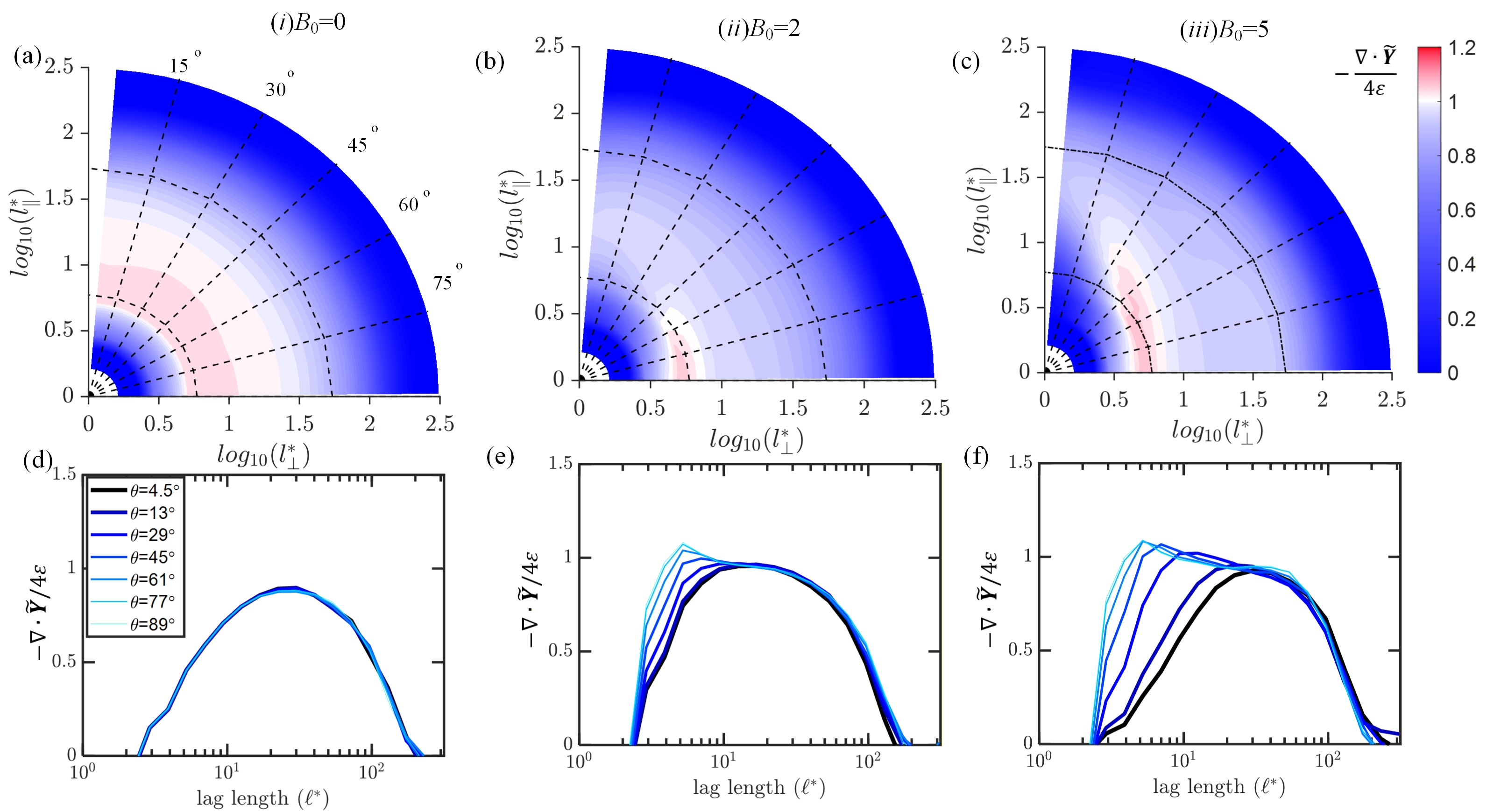}	
	\caption{Top row: image plots of normalized divergence of (azimuthally averaged) energy flux $ \wt{\vct{Y}} $.
    Bottom row: selected cuts at constant $\theta$ from the top row.
    Columns are for
    (i): $B_0 =0$; 
    (ii): $B_0 =2 $; 
    (iii): $B_0 = 5$. 
   Recall $\theta$ is the angle between the lag $\vct{\ell}$ and $\vct{B}_0 $.
    Black dashed arcs indicate inertial range boundaries, i.e., normalized lag lengths $\ell^*_\parallel, \ell^*_\perp \in [6,55]$, as identified using the direction-averaged form of the third-order law, i.e., method III,
    Eq.~\eqref{eq:3rd_order_sf_1D_lag}, with $-3(\overline{ Y_l^+}+\overline{ Y_l^-})/({8\varepsilon l})$ above a threshold, here 0.9. 
   } 	\label{fig:divY_contours}
\end{figure}

To directly illustrate the angular dependence of the energy transfer, Figure~\ref{fig:divY_contours} displays the divergence of the total energy flux vector,  
normalized by an appropriate multiple of the dissipation rate and azimuthally averaged;
  see Eq.~\eqref{eq:Yvector_components}.
Starting with the $B_0 = 5 $ simulations and comparing Figure~\ref{fig:divY_contours}f with Figure~\ref{fig:averaged_fixed_theta}a---which employs $\wt{Y_\ell}$ rather than the divergence of $\wt{\vct{Y}}$---we see that the 
main differences occur in the range of $\theta \approx[15^\circ,~45^\circ]$, in which method II in Eq.~\eqref{eq:3rd_order_sf_2D_lag} overestimates the true dissipation rate (albeit this observation may change for larger Reynolds number simulations with wider inertial ranges). 
For $ B_0 = 0 $, 
Figure~\ref{fig:divY_contours}d reveals that there is a range
over which all the curves collapse, indicating the expected angular independence, or isotropy. This range gets shorter with increasing $B_0$, however. 
To investigate these trends with $B_0$, in the Appendix we compare the results of this section with those from standard-dissipation cases.  That analysis verifies that our main conclusions are still valid when hyper-viscosity is employed.

\begin{figure}
	\centering
\includegraphics[width=0.85\linewidth,keepaspectratio]{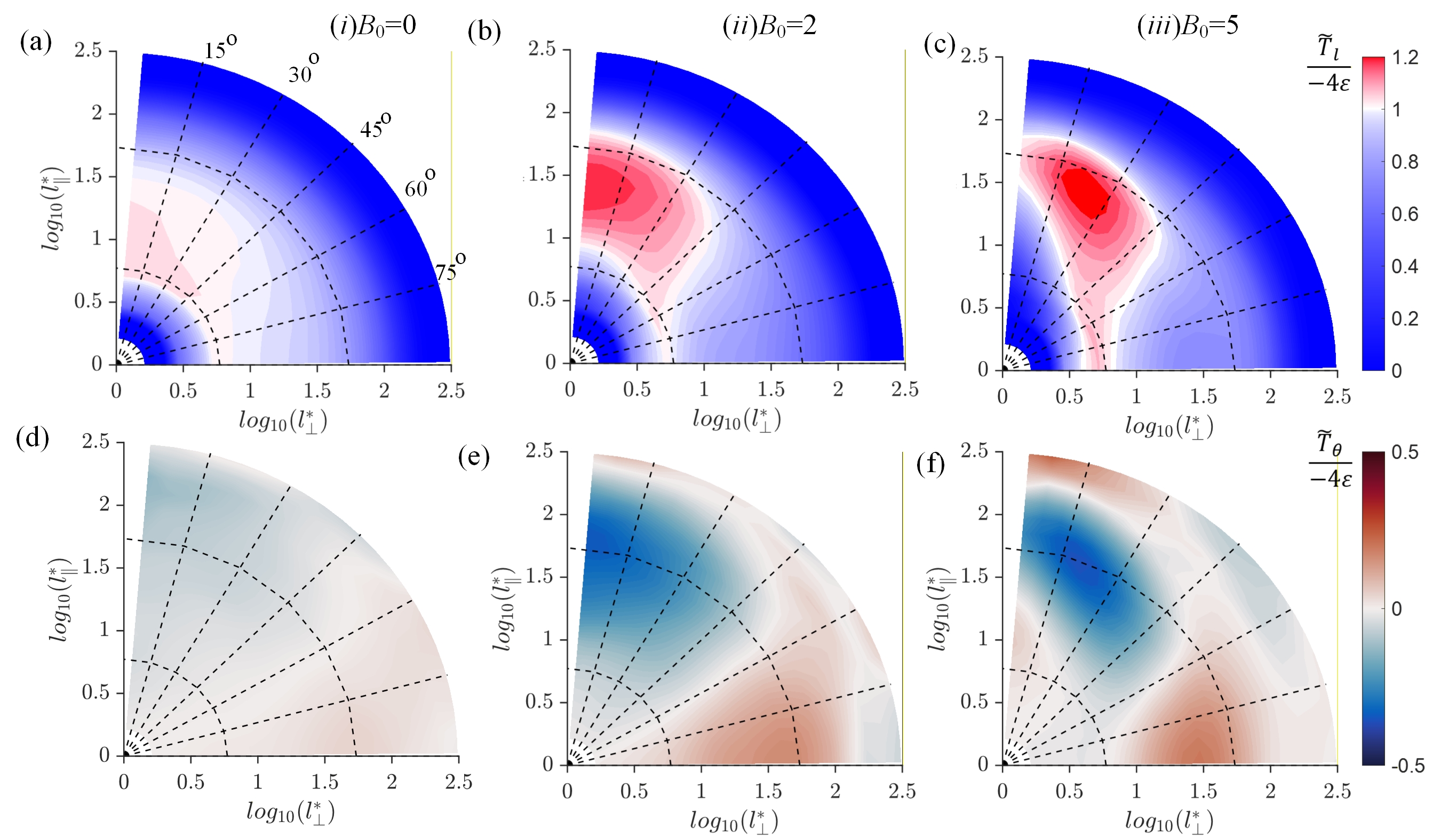}	\\[0.2cm]
 \caption{Image plots of the (top row) lag contribution $ \wt{T_\ell}$ 
 and (bottom row) polar contribution $ \wt{T_\theta} $ to the normalized divergence of the (azimuthally averaged) energy flux for 
    $B_0=0$ (panels a,d),  2 (b,e), and 5 (c,f). 
 Black dashed arcs indicate inertial range boundaries, i.e., normalized lag lengths $\ell^*_\parallel,\ell^*_\perp \in [6,55]$, as identified using the direction-averaged form of the third-order law, i.e., method III,
    Eq.~\eqref{eq:3rd_order_sf_3D_lag}, with $-3(\overline{ Y_\ell^+}+\overline{ Y_\ell^-})/({8\varepsilon \ell})$ above a threshold, here 0.9.
    Plotted quantities have been averaged over time and $\phi$.
 }
\label{fig:Divergence_components_SF3}
\end{figure}

Figure~\ref{fig:Divergence_components_SF3} shows the $ \wt{T}_\ell $ and $ \wt{T}_\theta$
additive contributions to the divergence of the energy flux vector; see Eq.~\eqref{eq:Yvector_components}. The $\wt{T_\phi}$ contribution (not shown here) is negligible, with its largest magnitude occurring for the strongest $B_0$ case. 
For the $\wt{T_\ell}$ contribution, which is single signed, the maximum value shifts slightly away from the parallel direction with increasing $B_0$. 
One also sees that the highest values with $ \wt{T}_\ell > 1$ are mainly located from $\theta \in[0, 45^\circ]$.
The $\wt{T_{\theta}}$ contribution can have positive and negative regions, and, for large enough $B_0$, there is a negative peak at smaller angles with $\theta \in [0, 45^\circ]$, and a positive peak at larger angles with $\theta \in [60^\circ, 90^\circ]$.  
When $ \wt{T_\theta} $ is added to 
 $ \wt{T}_\ell $,
 the negative peak region of the former offsets the `overlarge' values of the latter in the $\theta \in [0, 45^\circ]$ region, consistent with the differences seen between  Figure~\ref{fig:divY_contours}f and Figure~\ref{fig:averaged_fixed_theta}a. 
Interestingly, at inertial range scales,
  $ \wt{T_\theta} $
is always approximately zero around the $\theta = 60^\circ $ line. 
Referring to Figure~\ref{fig:averaged_fixed_theta}, we see this is the same angle for which $ \wt{Y_\ell}(\theta) \approx \overline{Y_\ell} $. 
 Since the $\theta$-average of $\wt{T_\theta}$, essentially
  $ \int_0^\pi \wt{T_\theta} \sin\theta \, \d\theta$, 
is zero, and $\wt{T_\theta}$ is continuous, the Mean Value Theorem of Integrals means that there exists at least one $\theta$  in the range $[0,~\pi]$ where $\wt{T_\theta} \sin\theta$ is zero. 
The empirical evidence from Figure~\ref{fig:Divergence_components_SF3} is that this occurs at $ \theta \approx 60^\circ$.

    \subsection{Possible general form of the energy flux}
    \label{sec:Similarity}

Having identified the importance of the (time and azimuthally
averaged) polar contribution to the divergence of the energy flux vector, 
  $ \wt{T_\theta} 
    = 
      \frac{1}{\ell \sin\theta} 
      \frac{\partial(\wt{Y_\theta} \sin\theta)}
           {\partial \theta} $, 
we now examine it in more detail in an effort to explain what is responsible for the speciality of $\theta = 60^\circ$.
\cite{podesta2007anisotropic} derived the general mathematical form of the energy flux vectors $\vct{Y}$ under the assumption that the turbulence is statistically axisymmetric for rotations about the direction of the mean magnetic field. 
His result, expressed in \emph{spherical} polar coordinates, is
\begin{align}
Y_{\ell,\text{Podesta}} &= \ell[A+B\cos^2\theta],
\label{eq:PGmodel_Yl}
  \\
Y_{\theta,\text{Podesta}} &= -B\ell \sin\theta\cos\theta,
\label{eq:PGmodel_Yth}
  \\
Y_{\phi,\text{Podesta}} &= C\ell \sin\theta,
\label{eq:PGmodel_Yphi}
  \\
Y_{\rho,\text{Podesta}} &= A\ell \sin\theta,
\label{eq:PGmodel_Yt}
\end{align}
where $A, B, C $ are functions of $\cos\theta$, that we may attempt to determine empirically.
Note that $A$ and $B$ are coupled by a differential equation, as can be seen by
substitution of the model into Eq.~\eqref{eq:div_3rd_order_law}
  \citep{podesta2007anisotropic}.
$Y_\rho$ is the component of the energy flux vector in the direction of $\vct{e_z} \times (\vct{\ell} \times \vct{e_z})$, which is parallel to the \emph{cylindrical} polar coordinate radial unit vector that lies in the $\vct{\ell}$--$\vct{B}_0$ plane. For simplicity, here we consider the zero-order solution in which both $A$ and $B$ are constants and $C=0$
  \citep{podesta2007anisotropic}. 

In a distinct approach, \cite{Galtier_2012}
assumed a power-law relation between correlation lengths along and transverse to the local mean magnetic field direction, and proposed a model for $\vct{Y}$. 
In cylindrical polar coordinates $(\rho, \phi, z)$ it is:
\begin{equation}
\vct{Y}_\text{Galtier} ( \rho,z) = -\frac{4\varepsilon} {3+a} 
 \left[\rho \vct{e_\rho} + (1+a)z \vct{e_z} \right],
\label{eq:PGmodel_cylinder}
\end{equation} 
where $a>0$ or $a<0$ represent convex or concave turbulence,
and $\vct{e}_\rho$ denotes the unit (cylindrical) radial vector. 
Rewriting this
in spherical coordinates we obtain
\begin{align}
Y_{\ell,\text{Galtier}} &= -\frac{4\varepsilon \ell} {3+a} (1+a\cos^2\theta),
\label{eq:Gmodel_Yl}
   \\
Y_{\theta,\text{Galtier}} &= \frac{4a\varepsilon \ell} {3+a} \sin\theta\cos\theta.
\label{eq:Gmodel_Yth}
   \\
Y_{\rho,\text{Galtier}} &= -\frac{4\varepsilon} {3+a}\ell \sin\theta,
\label{eq:Gmodel_Yt}
\end{align} 
We see that the zero-order model from \cite{podesta2007anisotropic} and the model from \cite{Galtier_2012} have the same polar angle dependence. 

  \begin{table} [tb]           
    \caption{Fitting parameters in the Podesta and Galtier models for $\vct{Y}$. }    
        \label{tab:fitting}
    \begin{center}
    \begin{tabular} { c | c c c } \hline
      Parameter & $B_0 = 0 $ & $ B_0 =2 $ & $ B_0 = 5 $ \\
       \hline
       $A$ & -2.53  & -1.93  &  -1.96 \\
       $B$ & -0.18 & -1.42  &  -1.60 \\
       \hline
       $a$ & 0     &  0.67  &  0.78  \\ \hline
    \end{tabular}
    \end{center}
  \end{table}

\begin{figure}[H]
	\centering
   \includegraphics[width=1\linewidth,keepaspectratio]{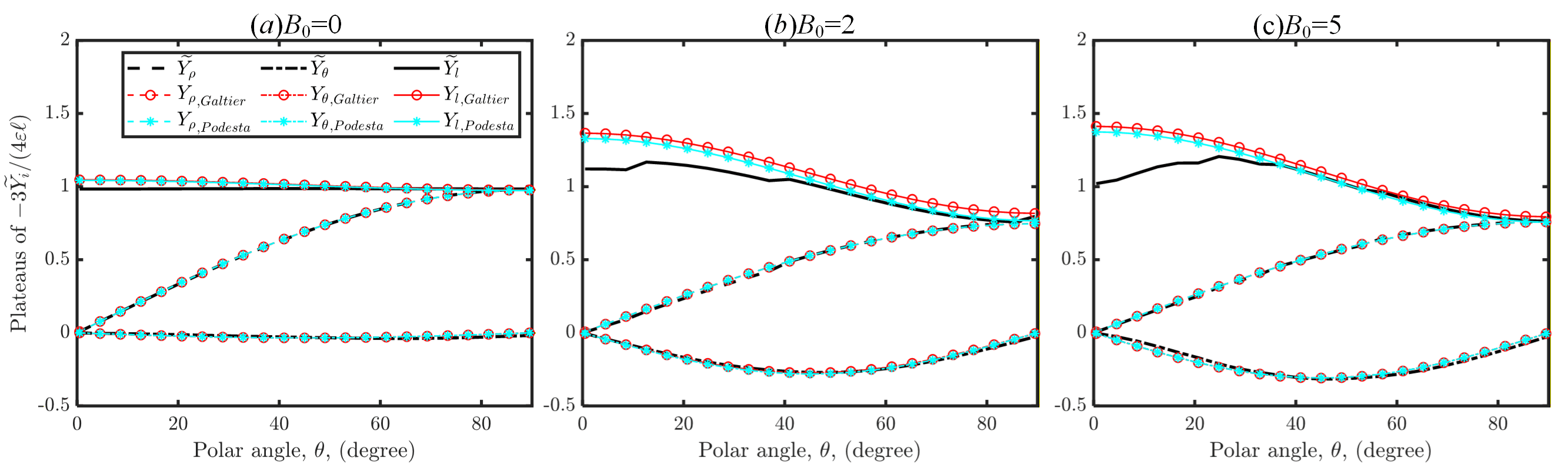}	
 \caption{Verification of the
  Podesta (blue, stars) and Galtier (red, circles)
 models for $\vct{Y}$.
 The dashed, dash-dotted, and solid curves are results for the $Y_\rho, Y_\ell, Y_\theta$ components, respectively. 
 Curves without symbols are obtained directly from DNS data.
   (a) $B_0= 0$, (b) $B_0 = 2$, and (c) $B_0 = 5 $. 
 }
\label{fig:PGmodel_verify}
\end{figure}

To help assess these models, we employ least-squares fitting of them to the plateau regions of the DNS-determined 
  $-3\wt{Y_\rho}/(4\varepsilon \ell)$ and $-3\wt{Y_{\theta}}/(4\varepsilon \ell)$;
that is, the fitting is only over the nominal inertial range scales.
The fitting parameters are listed in Table~\ref{tab:fitting} and the models based on them  are plotted in 
  Figure~\ref{fig:PGmodel_verify}.
One sees that the models for $Y_\theta$ and $Y_\rho$ agree well with our simulation results, while there are more observable differences for $Y_{\ell}$ (solid curves), especially for the $ B_0 = 5 $ case. 
In the future, one may also test the higher-order solution of the model from \cite{podesta2007anisotropic}, i.e., take into account the angular-dependent coefficients $A(\cos{\theta})$, $B(\cos{\theta})$ and $C(\cos{\theta})$.

\begin{figure}[H]
 \includegraphics[width=\linewidth]{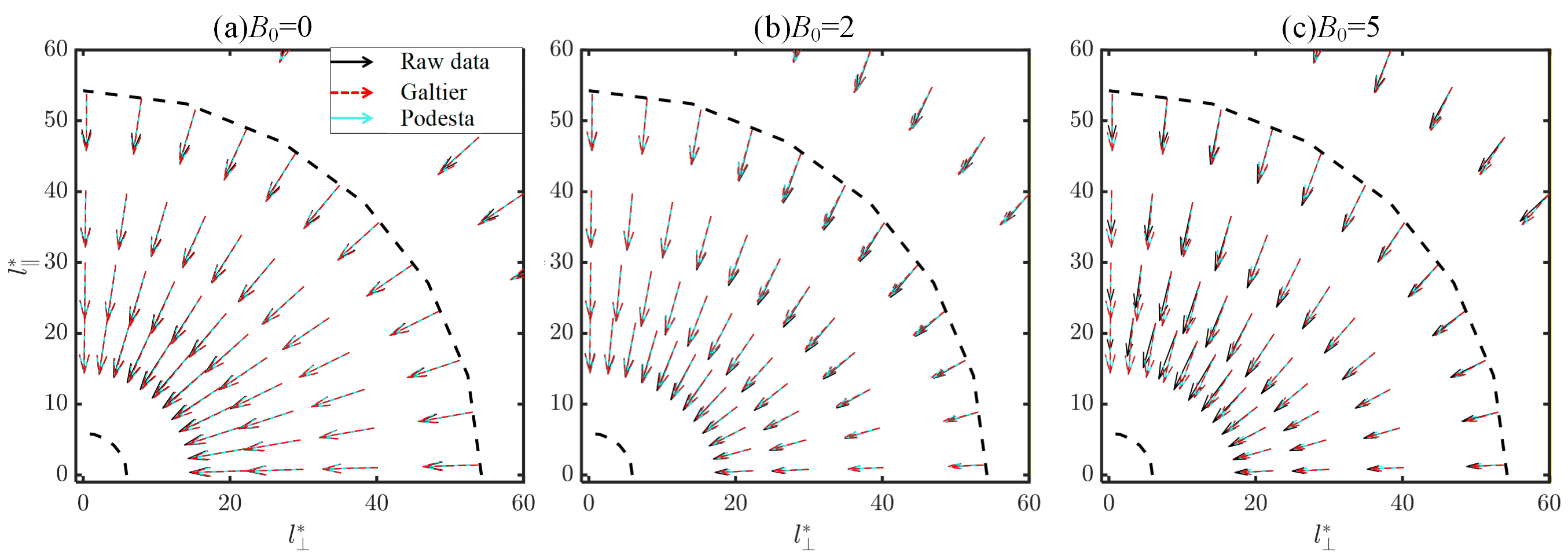}
 \caption{Vectors (in black) of the (azimuthally and time averaged) energy flux $ \wt{\vct{Y}} $ in parallel and perpendicular lag space, as obtained using DNS data for 
 (a) $B_0= 0$, (b) $B_0 = 2$, and (c) $B_0 = 5 $. 
 Vectors are normalized by the lag length, $l^*$. 
    Red dashed vectors are for the \cite{Galtier_2012} model, and light blue vectors for the \cite{podesta2007anisotropic} model. 
    Black dashed arcs indicate inertial range boundaries, i.e., normalised lag length $\ell^*_\parallel,\ell^*_\perp \in [6,55]$, as identified using the direction-averaged form of the third-order law, i.e., method III,
    Eq.~\eqref{eq:3rd_order_sf_3D_lag}, with $-3(\overline{ Y_\ell^+}+\overline{ Y_\ell^-})/({8\varepsilon \ell})$ above a threshold, here 0.9.
  }
\label{fig:Yvector01}
\end{figure}

Figure~\ref{fig:Yvector01} displays the vectors of the (time and azimuthally averaged) energy flux $ \wt{\vct{Y}} $ as a function of the parallel and perpendicular lag coordinates, for the $B_0= 0$, 2, and 5 simulations. Overall, we see that the energy flux vectors are well represented by the models of  \cite{podesta2007anisotropic} and \cite{Galtier_2012}. Some overestimation appears for the $B_0=5$ case, especially near the parallel direction and close to the dissipation range. 
For anisotropic cases, $B_0=5$, with a fixed length scale, the arrow length is the longest around $\theta \approx 30^\circ$, consistent with the trend of $\wt{Y_{\ell}}$ in Figure~\ref{fig:PGmodel_verify}. 

The model(s) for $Y_{\theta}$ can also provide an explanation for the speciality of the $60^\circ$ polar angle discussed in Sections~\ref{sec:observation} and \ref{sec:Eflux}. 
Substituting the Podesta model for $Y_{\theta} $, 
  Eq.~\eqref{eq:PGmodel_Yth}, 
into $T_{\theta}$ (see Eq.~\eqref{eq:Yvector_components})
gives
\begin{equation}
 {T_\theta} 
  = 
   \frac{1}{\ell \sin\theta}  
   \frac{\partial(\sin\theta \, {Y_\theta})} {\partial \theta}  
  = 
    - B(\cos2\theta + \cos^2\theta).
\label{eq:Pmodel_Tth}
\end{equation} 
Setting ${T_\theta}=0$, one finds $\theta = 54.7^\circ$. This supports our empirical results regarding the speciality of 
  $ \theta \approx 60^\circ $. 

    \subsection{Azimuthal angle dependence}
    \label{sec:azimuthal}

In previous works, the turbulence has often been assumed to be statistically axisymmetric about the direction of the mean magnetic field (also called azimuthally symmetric or cylindrically symmetric). Having shown the speciality of $60^\circ$ polar angle with the azimuthal average, in this subsection we go beyond the axisymmetric model and demonstrate the azimuthal dependence of the third-order structure function. 

\begin{figure}[H]
	\centering
\includegraphics[width=\linewidth,keepaspectratio]{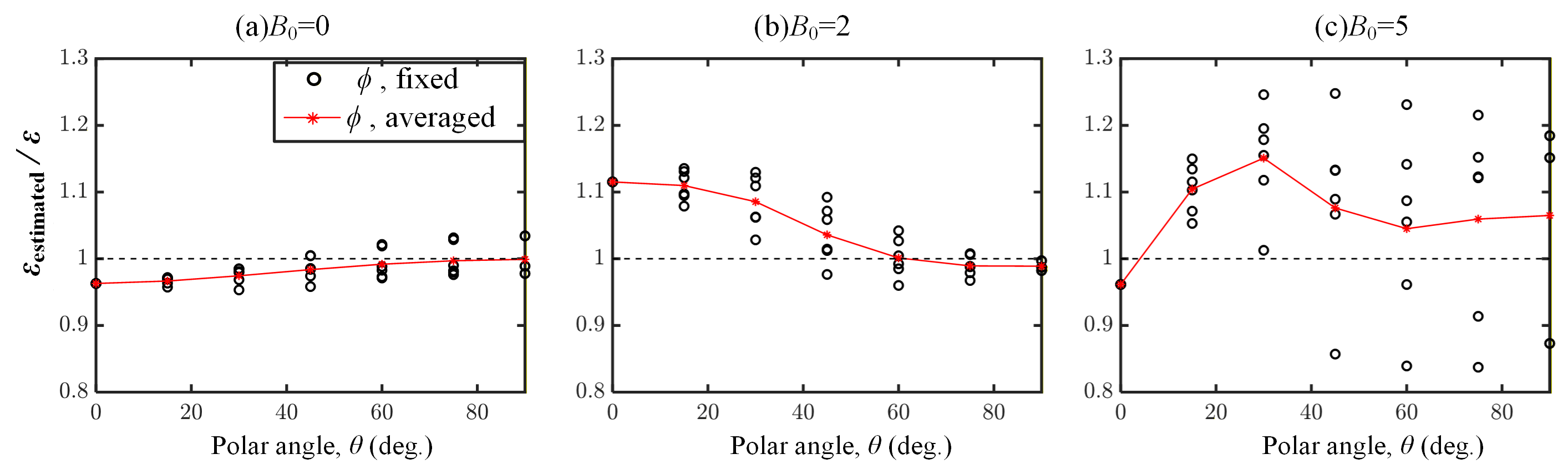} 
	\caption{Dissipation rates estimated from normalized longitudinal third-order structure functions for various $\theta$ and $\phi$, as obtained using method I, Eq.~\eqref{eq:3rd_order_sf_3D_lag}. Values are normalized to the total dissipation rate, $ \varepsilon $. 
    The solid red line with stars represents the azimuthally averaged profiles, as obtained using method II in Eq.~\eqref{eq:3rd_order_sf_2D_lag}. 
    Data has been time averaged.
    (a): $ B_0 = 0 $;
    (b): $ B_0 = 2 $;  
    (c): $ B_0 = 5$.
    }
	\label{fig:Singlephi_hyper2e7_B2_5}
\end{figure}

Figure \ref{fig:Singlephi_hyper2e7_B2_5} displays the distribution of the estimated dissipation rates at different $\theta$ and $\phi$, marked with dark circles, for three values of $B_0$.  
The red curves represent the azimuthal averages and
we see that the $60^\circ$ polar angle gives the most accurate result in the anisotropic cases ($B_0=2$, 5),
with the values $ \theta > 60^\circ $ 
also being quite accurate.
It is evident that the distributions of estimated dissipation rates at fixed $\theta$ and varying $\phi$ are more scattered for larger $B_0$. 
For $B_0 = 2$ and 5, the maximum estimated cascade rate can depart from the actual value by 10$\%$ and $25\%$, respectively. 
We expect this departure to be even greater at larger $B_0$. 
Nonetheless, as shown by the red line with stars in Figure \ref{fig:Singlephi_hyper2e7_B2_5}(b,c), after averaging over azimuth, the maximum departures from unity reduce to 3$\%$ and 15$\%$ for $B_0=2$ and 5, respectively. 
From a practical perspective, azimuthal coverage generally accompanies polar coverage in spacecraft observations. 
Moreover, given that it is almost impossible that all intervals will have the same $\phi$ when a large number of intervals are used, it may be feasible to achieve sufficient coverage over $\phi$. 
More quantitative assessment on the $\phi$ dependence shall be done in our companion paper.

Theoretically, assuming axisymmetry holds, when the number of time snapshots employed is sufficient to compute a stable average, the averaged structure function should be independent of azimuthal angle. Thus, when we describe energy transfer as isotropic in the perpendicular plane, we are referring to statistically averaged transfer. When sampling and averaging are limited, 
e.g., a small number of snapshots (as in DNS) or sampling directions (as in observations), then a residual dependence on the azimuthal directions might persist in the estimates. This dependence could be associated with local spatial and temporal fluctuations caused by large-scale structures in the perpendicular plane at large $B_0$, as observed by \cite{zikanov1998direct} for example, 
and also seen in our Figure~\ref{fig:structures}f.

\subsection{Verification with virtual space observation}
    \label{sec:spacecraft}
To verify our observation on the speciality of $60^\circ$ polar angle, we also perform virtual spacecraft measurements. That is, we employ four spacecraft in tetrahedral configurations to fly through numerically generated turbulent fields, mimicking satellite (e.g, \emph{MMS} and \emph{Cluster} mission) flights through solar wind and magnetosheath turbulence.
The relative positions of the virtual spacecraft are scaled to fit within the simulation domain and the interspacecraft separation is set to be 37 times the 
(velocity) Kolmogorov length scale, i.e., in the inertial range. 
The trajectories of the virtual spacecraft are parallel lines with specified polar and azimuthal angles. Due to the periodic boundary conditions of the simulations, the spacecraft trajectories cross the simulation box several times; 
 see Figure~2a in \cite{Pecora2023ApJ}. 

\begin{figure}[H]
	\centering
	\includegraphics[width=0.8\linewidth,keepaspectratio]{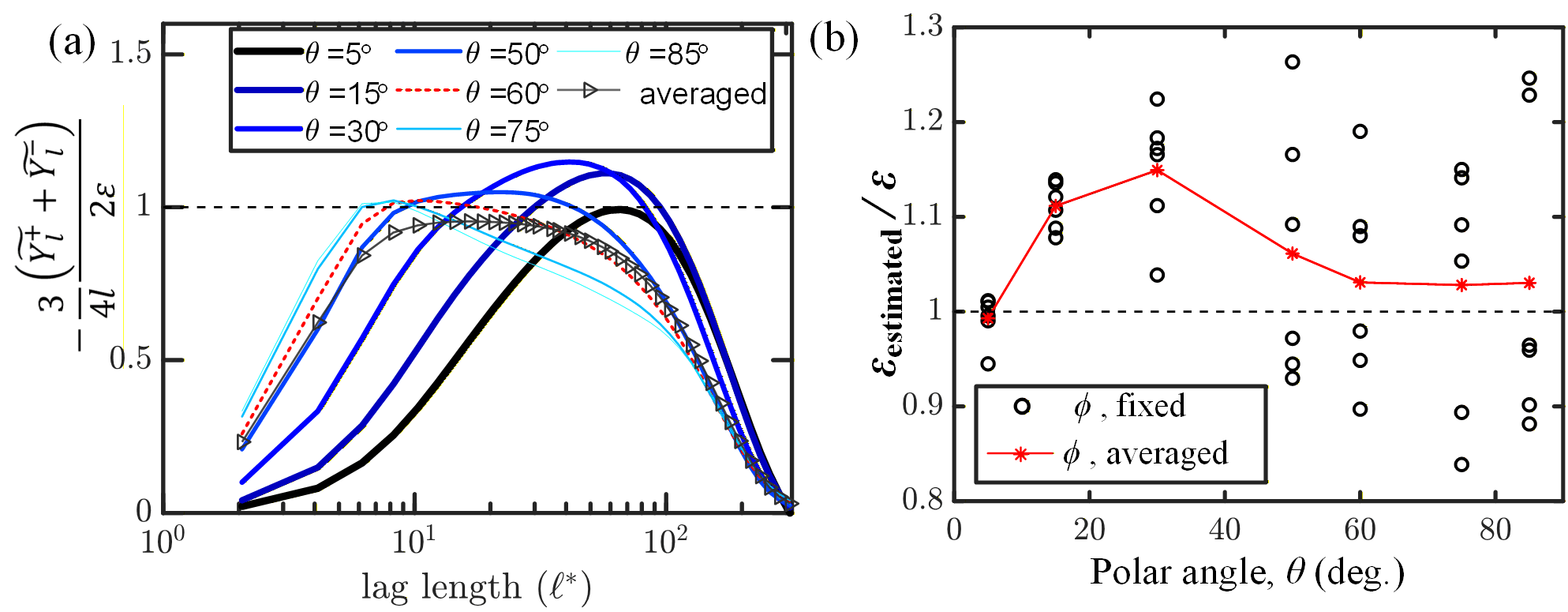}	
	\caption{Results from virtual spacecraft measurements for the $B_0=5$ simulation at indicated values of $\theta$:  
    (a) Azimuthal average. 
    The solid line with triangles is the direction-averaged profile. 
    (b) Cascade rates estimated from normalized longitudinal third-order structure functions for various $\theta$ and $\phi$. Values are normalized to the total dissipation rate, $ \varepsilon $. 
    The solid red line with (red) stars represents the azimuthally averaged profiles. 
    Data has been time averaged with the same snapshots as figure~\ref{fig:Singlephi_hyper2e7_B2_5}(c).
    } 
	\label{fig:Vspace}
\end{figure}

The MHD simulation we are using here is the $B_0=5$ case. 
Temporal sampling is as indicated in Table~\ref{table:setup}, with 75 snapshots over about 75 large-eddy turnover times. The sampling polar angles are 
 $ \theta = [5^\circ, 15^\circ, 30^\circ, 50^\circ, 60^\circ, 75^\circ, 85^\circ]$,
and azimuthal angles 
 $ \phi = [0^\circ : 60^\circ : 300^\circ] $.
The mean number of data points
for each spacecraft along one trajectory, i.e., one direction, is about $ 1.6 \times 10^5  $  per snapshot. 
For each trajectory, the third-order structure function is calculated for 40 lag lengths spanning 2 to 314 times the (velocity) Kolmogorov length scale. 
The longitudinal third-order structure function for a fixed polar and azimuthal angle, ${ Y_\ell^{\pm}}{(\ell, \theta_i, \phi_j)}$, is averaged from four spacecraft,
\begin{equation}
{ Y_\ell^{\pm}}{(\ell, \theta_i, \phi_j)} 
 = 
\frac{\sum_{k=1}^{N_k=4} \sum_{t=1}^{N_t} Y_\ell^{\pm}(\ell, \theta_i,\phi_j,t,k) }{N_kN_t},
\label{eq:VS_average}
\end{equation} 
where $N_k$ and $N_t$ denote the number of virtual spacecraft and snapshots, respectively.
Comparing Figure~\ref{fig:Vspace}a with Figure~\ref{fig:averaged_fixed_theta}a, we see that the result at $\theta = 60^\circ$ is still the best approximation to the profile with direction-averaged method, especially for the plateau. 
The results from virtual spacecraft at different azimuthal angles, 
 shown in Figure~\ref{fig:Vspace}b, 
are in general the same as the DNS results in Figure~\ref{fig:Singlephi_hyper2e7_B2_5}c.  
Indeed, Figures~\ref{fig:Singlephi_hyper2e7_B2_5} and~\ref{fig:Vspace} both indicate that there is a strong dependence of the estimated dissipation rate on the azimuthal angle $\phi$. 
When employing spacecraft observations, the azimuthal average could be performed using different measurement intervals, similar to the polar averaging reported on in  \cite{Osman2011Anisotropic} and \cite{Bandyopadhyay_2018}. 
Based on the results presented here, we suggest that the main objective should be selection of intervals with $\theta \approx 60^\circ $, and a range of $ \phi $ values (to average over).

    \section{Conclusions and discussions}
    \label{sec:conclusions}

Anisotropy, a typical property of MHD-scale fluctuations in the solar wind, challenges the applicability of the third-order law with the isotropic assumption. The literature \citep[e.g.,][]{Osman_2011heating,Bandyopadhyay_2018,wang2022strategies} 
presents a direction-averaged form to resolve this challenge, although it requires full angle coverage and a large number of space datasets in solar wind measurement.
Regarding various angle averages of the MHD longitudinal third-order moment, $Y_\ell$, in this work we have found that:

\begin{enumerate}
 \item 
 The azimuthally-averaged third-order structure function at $\theta \approx 60^\circ$ can predict the (full) direction-averaged results, that is, 
    $ \wt{Y_\ell} (\theta \approx 60^\circ) \approx  \overline{Y_\ell}$. 
The agreement holds for mean magnetic field strengths up to at least $B_0 = 5 $. 
However, the deviation increases with increasing $B_0$, but is still within 10$\%$ differences at $B_0 = 5 $. 

  \item
This speciality of $60^\circ$ polar angle relative to the mean magnetic field direction can be explained by considering the divergence of energy flux vector, $\nabla_\ell \cdot \vct{Y}$.  
In the inertial range, the $ \wt{Y_\theta} $ contribution to the divergence, 
namely $ \wt{T_\theta}$,  integrates (wrt $\theta$) to zero. 
Based on the Mean Value Theorem of Integrals, this suggests that there exists at least one $\theta$ in the range $[0,~\pi]$ where $\wt{T_\theta}$ is zero. 
Furthermore, we find that the $\theta$-averaged $ T_\theta \sin{\theta}$, is negative over $[0^\circ,\; 60^\circ]$ and positive over $[60^\circ,\; 90^\circ]$, which gives rise to this special angle  $60^\circ$.

\item 
In the theory of the anisotropic form of the third-order moments and their relationship to the energy dissipation rates, the zero-order solution for the model from \cite{podesta2007anisotropic} and the model from \cite{Galtier_2012} are assessed, and support our observation on the speciality of $60^\circ$ polar angle.

\item 
The dependence on azimuthal angles $\phi$ is assessed. We find that the distributions of estimated dissipation rates at fixed $\theta$ and varying $\phi$ are more scattered for larger $B_0$. At least in part, we expect that this is related to the effects of insufficient sampling of large-scale structures present in the available data. 
However, there are sound theoretical reasons to expect activity in the perpendicular planes to be isotropic.

  \item 
Based on (tetrahedral) virtual spacecraft measurements, 
analogous to the \emph{MMS} and \emph{Cluster} mission, 
the speciality of $60^\circ$ polar angle and azimuthal dependence are further verified. These results provide guidance for real spacecraft measurements.

\end{enumerate}

These findings can assist with determining accurate estimates of the energy dissipation rates in the solar wind using the typically directionally-limited observations that are available. 

\begin{acknowledgments}
This work was supported by Xiangtan University Start-up research grant (Grant No.KZ0812269); Y. Y is partially supported by the University of Delaware General University Research Program grant. M.W is partially supported by NSFC (Grant Nos. 12225204 and 11902138); Department of Science and Technology of Guangdong Province (Grant Nos. 2019B21203001 and 2020B1212030001); the Shenzhen Science and Technology Programme (Grant No. KQTD20180411143441009).
The assistance of resources and services from the Centre for Computational Science and Engineering of Southern University of Science and Technology is also acknowledged. The authors thank for helpful discussions with Yuchen Ye.
\end{acknowledgments}

\begin{contribution}
All authors contributed equally to the Terra Mater collaboration.
\end{contribution}

\appendix
\section*{Results from standard Laplacian dissipation simulations}
\begin{figure}[H]
\centering
   \includegraphics[width=0.8\linewidth,keepaspectratio]
{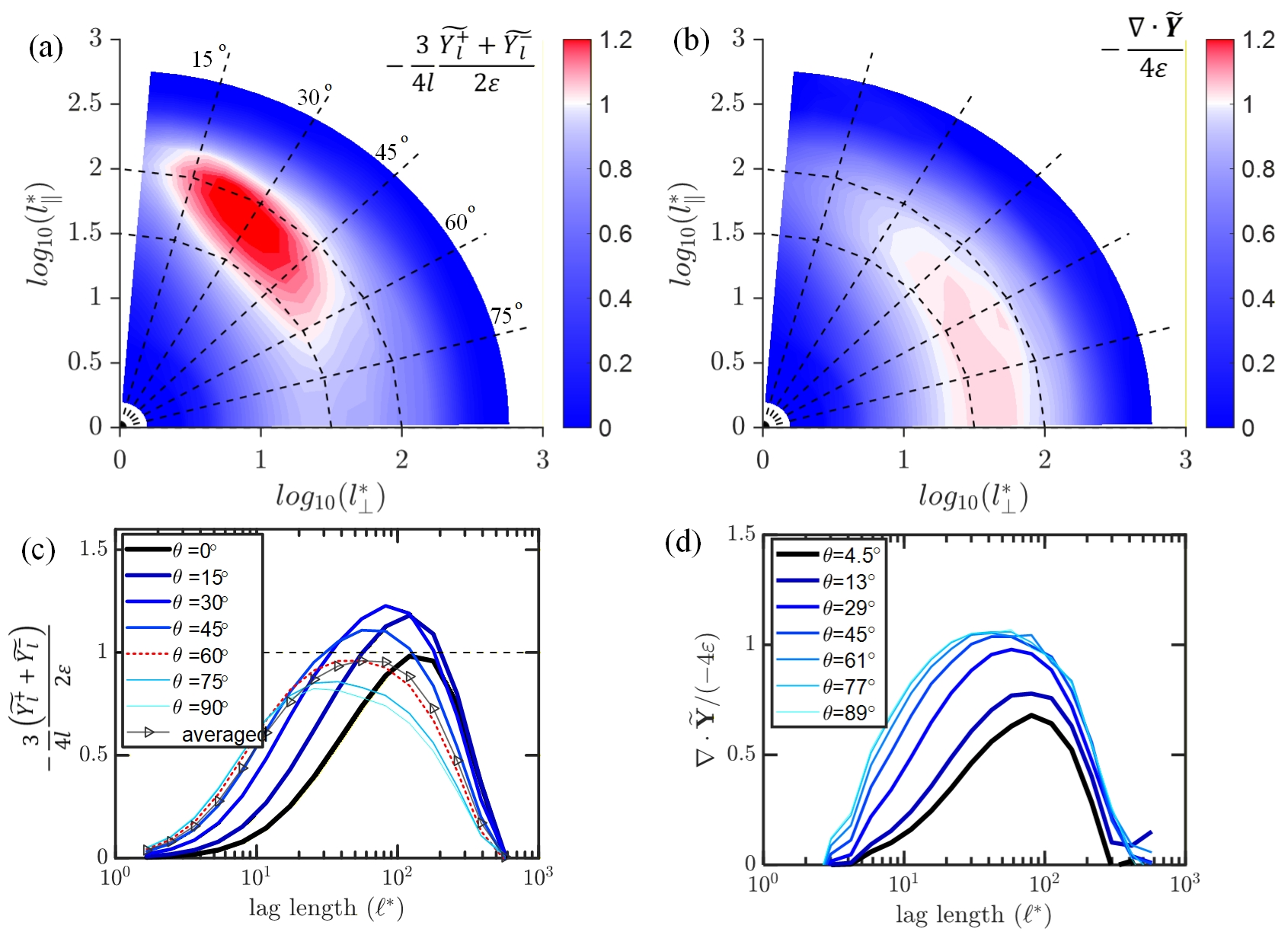}	\\[0.2cm]
 \caption{Laplacian dissipation results for a $B_0 = 5$ simulation: 
 (a,c) azimuthally and time averaged and normalized longitudinal third-order structure functions ($ \wt{Y_\ell} $); see Eq.~\eqref{eq:3rd_order_sf_2D_lag}.
 The angle between the lag vector and $ \vct{B}_0 $ is $\theta $. 
(b,d) Divergence of the (azimuthally averaged) energy flux vector.
  Black dashed arcs represent the inertial range boundaries, i.e., lag length $\ell^*_\parallel,\ell^*_\perp \in [32,100]$, as identified using the direction-averaged third-order law, i.e., method III, 
 Eq.~\eqref{eq:3rd_order_sf_1D_lag}, with
 $ -3 (\overline{ Y_\ell^+} + \overline{ Y_\ell^-}) / ({8\varepsilon \ell}) $ 
 above a threshold, here 0.9.
 Curves shown in the bottom row are obtained from $\theta = $constant cuts through the images in the top row.
The setup for Laplacian dissipation with 1024$^3$-grids is discussed in \cite{jiang2023energy}.
 }
 \label{fig:Yl_divY_Norm_B5_V5}
\end{figure}

Here we compare results obtained using standard Laplacian dissipation with the hyper-dissipation results from the main body of the paper (e.g., Figures~\ref{fig:averaged_fixed_theta} and~\ref{fig:divY_contours}), focusing on $B_0 = 5$ simulations. In Figure~\ref{fig:Yl_divY_Norm_B5_V5}, the lefthand column displays the (azimuthally and time averaged) normalised longitudinal third-order structure functions for a Laplacian dissipation case, as an image plot in the $\ell_\perp$--$\ell_\parallel$ plane and as cuts through this image at fixed $\theta$.
Comparing these panels with those of Figure~\ref{fig:averaged_fixed_theta}, one sees that for both types of dissipation, the plateaus associated with the structure functions at $ \theta = 60^\circ$ are closest to the fully ($\theta$ and $\phi$) angle-averaged results. Consequently, these plateaus levels predict the actual energy dissipation rate reasonably accurately.

Figure~\ref{fig:Yl_divY_Norm_B5_V5}(b,d)  shows the divergence of the energy flux for this same standard Laplacian dissipation run with $B_0 = 5$. 
Compared with the hyper-viscous case (see Figure~\ref{fig:divY_contours}), the location of the peak dissipation rate shifts to larger scales for this standard viscous case, and the peak value associated with the parallel direction decreases. 
This indicates that, for the hyper-dissipative case, the energy transfer in the parallel direction is enhanced and the strength of perpendicular small-scale turbulence structures is increased.
In Figure~\ref{fig:divY_contours}, one can argue that the small-scale anisotropy increases with $B_0$, but there is always an $\ell$ range where there is isotropy wrt all $\theta$. For the standard dissipation case, all scales are anisotropic (although only weakly at the larger scales). We suspect that the difference between standard/hyper-dissipation cases here might be because in the standard Laplacian case the inertial range is not wide enough, especially for the smaller polar angles.

\bibliography{Reference}{}

@ARTICLE{AdhikariEA25-SWvisc,
doi = {10.3847/1538-4357/ae101d},
url = {https://doi.org/10.3847/1538-4357/ae101d},
year = {2025},
month = {oct},
publisher = {The American Astronomical Society},
volume = {993},
number = {1},
pages = {60},
author = {Adhikari, Laxman and Giri, Ashutosh and Karki, Monika and Baruwal, Prashant and Baruwal, Prashrit and Khondoker Shikha, Rubaiya and Mainali, Bigyan and Kenno, Dessalegn and Tasnim, Ismita and Zank, Gary P. and Wang, Bingbing and Treville, Lucien and Ghimire, Sagar},
title = {Reynolds Number and Viscosity in the Solar Wind},
journal = {The Astrophysical Journal},
}

@ARTICLE{AdhikariEA25-effvisc,
doi = {10.3847/1538-4357/ae08b4},
url = {https://doi.org/10.3847/1538-4357/ae08b4},
year = {2025},
month = {oct},
publisher = {The American Astronomical Society},
volume = {992},
number = {2},
pages = {180},
author = {Adhikari, Subash and González, Carlos A. and Yang, Yan and Oughton, Sean and Pecora, Francesco and Bandyopadhyay, Riddhi and Matthaeus, William H.},
title = {Estimation of Effective Viscosity to Quantify Collisional Behavior in Collisionless Plasma},
journal = {The Astrophysical Journal},
}

@article{Bandyopadhyay_2018,
doi = {10.3847/1538-4357/aade04},
url = {https://doi.org/10.3847/1538-4357/aade04},
year = {2018},
month = {oct},
publisher = {The American Astronomical Society},
volume = {866},
number = {2},
pages = {106},
author = {Bandyopadhyay, Riddhi and Chasapis, A. and Chhiber, R. and Parashar, T. N. and Matthaeus, W. H. and Shay, M. A. and Maruca, B. A. and Burch, J. L. and Moore, T. E. and Pollock, C. J. and Giles, B. L. and Paterson, W. R. and Dorelli, J. and Gershman, D. J. and Torbert, R. B. and Russell, C. T. and Strangeway, R. J.},
title = {Incompressive Energy Transfer in the Earth’s Magnetosheath: Magnetospheric Multiscale Observations},
journal = {The Astrophysical Journal},
}

@BOOK{      BatchelorTHT,
    Bibkey   = {BatchelorTHT},
    AUTHOR   = {Batchelor, G. K.},
    TITLE    = {The Theory of Homogeneous Turbulence},
    YEAR     = {1970},
    Publisher = {Cambridge University Press},
    Address  = {Cambridge, UK},
    Comments = {Turbulence text; reissue of 1953 classic}
}

@ARTICLE{BrunoCarboneLRSP13,
doi = {10.12942/lrsp-2013-2},
       author = {{Bruno}, R. and {Carbone}, V.},
        title = "{The Solar Wind as a Turbulence Laboratory}",
      journal = {Living Reviews in Solar Physics},
         year = "2013",
        month = "May",
       volume = {10},
       number = {1},
          eid = {2},
        pages = {2}
}

@Article{burch2016MMS, author={Burch, J. L. and Moore, T. E. and Torbert, R. B. and Giles, B. L.}, title={Magnetospheric Multiscale Overview and Science Objectives}, journal={Space Science Reviews}, year={2016}, month={3}, day={01}, volume={199}, number={1}, pages={5-21}, doi ={10.1007/s11214-015-0164-9}}

@book{canuto2007spectral,
  title={Spectral methods: evolution to complex geometries and applications to fluid dynamics},
  author={Canuto, Claudio and Quarteroni, Alfio and Hussaini, M Yousuff and Zang Jr, Thomas A},
  year={2007},
  publisher={Springer}
}

@article{coleman1968turbulence,
  title={Turbulence, viscosity, and dissipation in the solar-wind plasma},
  author={Coleman, J.r. and Paul, J.},
  journal={Astrophysical Journal},
  volume={153},
  pages={371},
  year={1968},
doi={10.1086/149674}
}

@article{Karman1938statistical,
    author = {de Karman, Theodore and Howarth, Leslie},
    title = {On the Statistical Theory of Isotropic Turbulence},
    journal = {Proceedings of the Royal Society of London. A. Mathematical and Physical Sciences},
    volume = {164},
    number = {917},
    pages = {192-215},
    year = {1938},
    month = {01},
    issn = {0080-4630},
    doi = {10.1098/rspa.1938.0013},
    url = {https://doi.org/10.1098/rspa.1938.0013},
    eprint = {https://royalsocietypublishing.org/rspa/article-pdf/164/917/192/35061/rspa.1938.0013.pdf},
}

@Article{escoubet2001cluster, AUTHOR = {Escoubet, C. P. and Fehringer, M. and Goldstein, M.}, TITLE = {IntroductionThe Cluster mission}, JOURNAL = {Annales Geophysicae}, VOLUME = {19}, YEAR = {2001}, NUMBER = {10/12}, PAGES = {1197--1200}, doi={10.5194/angeo-19-1197-2001}}

@BOOK{FrischBook95,
  author = {Frisch, U.},
    title = "{Turbulence. The legacy of A. N. Kolmogorov.}",
booktitle = {Turbulence.~The legacy of A.~N.~Kolmogorov., by U. Frisch, Cambridge, UK: Cambridge University Press, 1995},
    publisher={Cambridge university press},
     year = 1995,
    month = nov
}

@article{Galtier_2012,
doi = {10.1088/0004-637X/746/2/184},
url = {https://doi.org/10.1088/0004-637X/746/2/184},
year = {2012},
month = {feb},
publisher = {The American Astronomical Society},
volume = {746},
number = {2},
pages = {184},
author = {Galtier, Sébastien},
title = {Kolmogorov vectorial law for solar wind turbulence},
journal = {The Astrophysical Journal},
}

@book{gottlieb1977numerical,
  title={Numerical analysis of spectral methods: theory and applications},
  author={Gottlieb, David and Orszag, Steven A},
  year={1977},
  publisher={SIAM}
}

@article{horbury2008anisotropic,
  title={Anisotropic scaling of magnetohydrodynamic turbulence},
  author={Horbury, Timothy S and Forman, Miriam and Oughton, Sean},
  journal={Physical Review Letters},
  volume={101},
  number={17},
  pages={175005},
  year={2008},
  publisher={APS},
doi={10.1103/PhysRevLett.101.175005}
}

@article{jiang2023energy,
  title={Energy transfer and third-order law in forced anisotropic magneto-hydrodynamic turbulence with hyper-viscosity},
  author={Jiang, B. and Li, C. and Yang, Y. and Zhou, K. and Matthaeus, W.H. and Wan, M.},
  journal={Journal of Fluid Mechanics},
  volume={974},
  pages={A20},
  year={2023},
  publisher={Cambridge University Press},
DOI={10.1017/jfm.2023.743}
}

@article{jokipii1970interplanetary,
  title={Interplanetary scintillations and the structure of solar-wind fluctuations},
  author={Jokipii, J.R. and Hollweg, Joseph V.},
  journal={Astrophysical Journal},
  volume={160},
  pages={745},
  year={1970},
doi={10.1086/150466}
}

@article{luo2010observations,
  title={Observations of anisotropic scaling of solar wind turbulence},
  author={Luo, QY and Wu, DJ},
  journal={The Astrophysical Journal Letters},
  volume={714},
  number={1},
  pages={L138},
  year={2010},
  publisher={IOP Publishing},
doi={10.1088/2041-8205/714/1/L138}
}

@article{macbride2008turbulent,
  title={The turbulent cascade at 1 AU: energy transfer and the third-order scaling for MHD},
  author={MacBride, B.T. and Smith, C.W. and Forman, M.A.},
  journal={Astrophysical Journal},
  volume={679},
  number={2},
  pages={1644},
  year={2008},
  publisher={IOP Publishing},
doi={10.1086/529575}
}

@INPROCEEDINGS{marcucci2024PO,        author = {{Marcucci}, Maria Federica and {Retin{\`o}}, Alessandro},         title = "{The ESA M7 candidate mission Plasma Observatory.}",     booktitle = {EGU General Assembly Conference Abstracts},          year = 2024,        series = {EGU General Assembly Conference Abstracts},         month = apr,           eid = {11903},         pages = {11903}, doi={ 
10.5194/egusphere-egu24-11903} }

@article{Marino2023,
author = {Marino, Raffaele and Sorriso-Valvo, Luca},
issn = {03701573},
journal = {Physics Reports},
pages = {1--144},
publisher = {Elsevier B.V.},
title = {{Scaling laws for the energy transfer in space plasma turbulence}},
volume = {1006},
year = {2023},
doi={10.1016/j.physrep.2022.12.001}
}

@article{matthaeus1996anisotropic,
  title={Anisotropic three-dimensional MHD turbulence},
  author={Matthaeus, W.H. and Ghosh, S. and Oughton, S. and Roberts, D.A.},
  journal={Journal of Geophysical Research: Space Physics},
  volume={101},
  number={A4},
  pages={7619--7629},
  year={1996},
  publisher={Wiley Online Library},
doi={10.1029/95JA03830}
}

@article{matthaeus1982measurement,
  title={Measurement of the rugged invariants of magnetohydrodynamic turbulence in the solar wind},
  author={Matthaeus, W.H. and Goldstein, M.L.},
  journal={Journal of Geophysical Research: Space Physics},
  volume={87},
  number={A8},
  pages={6011--6028},
  year={1982},
  publisher={Wiley Online Library},
doi={10.1029/ja087ia08p06011}
}

@misc{MatthaeusEA19-whitepaper, 
    author = {Matthaeus, W. H. and Bandyopadhyay, R. and Brown, M. R. and Borovsky, J. and Carbone, V. 
	and Caprioli, D. and Chasapis, A. and Chhiber, R. and Dasso, S. and Dmitruk, P. and Del Zanna, L. 
	and Dmitruk, P. A. and Franci, Luca and Gary, S. P. and Goldstein, M. L. and Gomez, D. and Greco, A. 
	and Horbury, T. S. and Ji, Hantao and Kasper, J. C. and Klein, K. G. and Landi, S. and Li, Hui and 
	Malara, F. and Maruca, B. A. and Mininni, P. and Oughton, Sean and Papini, E. and Parashar, T. N. 
	and Petrosyan, Arakel and Pouquet, Annick and Retino, A. and Roberts, Owen and Ruffolo, David and 
	Servidio, Sergio and Spence, Harlan and Smith, C. W. and Stawarz, J. E. and TenBarge, Jason and Vasquez1, 
	B. J. and Vaivads, Andris and Valentini, F. and Velli, Marco and Verdini, A. 
	and Verscharen, Daniel and Whittlesey, Phyllis and Wicks, Robert and Bruno, R. and Zimbardo, G.},
    title = {[Plasma 2020 Decadal] The essential role of multi-point measurements in turbulence investigations: the solar wind beyond single scale and beyond the Taylor Hypothesis},
    publisher = {arXiv},
    year = {2019},
    copyright = {arXiv.org perpetual, non-exclusive license},
doi={10.48550/arXiv.2211.12676}
}

@BOOK{MoninYaglom,
    AUTHOR   = {Monin, A.S. and Yaglom, A.M.},
    TITLE    = {Statistical Fluid Mechanics, Vols 1 and 2},
    YEAR     = {1971, 1975},
    Publisher= {MIT Press},
    Address  = {Cambridge, Mass.},
}

@article{NieTanveer99,
  title={A note on third--order structure functions in turbulence},
  author={Nie, Q. and Tanveer, S.},
  journal={Proceedings of the Royal Society of London. Series A: Mathematical, Physical and Engineering Sciences},
  volume={455},
  number={1985},
  pages={1615--1635},
  year={1999},
  publisher={The Royal Society},
doi={10.1098/rspa.1999.0374}
}

@article {Orszag1971,
      author = "Steven A.  Orszag",
      title = "On the Elimination of Aliasing in Finite-Difference Schemes by Filtering High-Wavenumber Components",
      journal = "Journal of Atmospheric Sciences",
      year = "1971",
      publisher = "American Meteorological Society",
      address = "Boston MA, USA",
      volume = "28",
      number = "6",
      pages=      "1074 - 1074",
doi={https://doi.org/10.1175/1520-0469(1971)028<1074:OTEOAI>2.0.CO;2}
}

@article{Orszag1972,
author = {Orszag, Steven A.},
title = {Comparison of Pseudospectral and Spectral Approximation},
journal = {Studies in Applied Mathematics},
volume = {51},
number = {3},
pages = {253-259},
year = {1972},
doi={10.1002/sapm1972513253}
}

@article{Orszag_Tang_1979, title={Small-scale structure of two-dimensional magnetohydrodynamic turbulence}, volume={90}, number={1}, journal={Journal of Fluid Mechanics}, author={Orszag, Steven A. and Tang, Cha-Mei}, year={1979}, pages={129–143},doi={10.1017/S002211207900210X}
}

@article{Osman_2011heating,
  title={Evidence for inhomogeneous heating in the solar wind},
  author={Osman, K.T. and Matthaeus, W.H. and Greco, A. and Servidio, S.},
  journal={Astrophysical Journal Letters},
  volume={727},
  number={1},
  pages={L11},
  year={2010},
  publisher={IOP Publishing},
doi={10.1088/2041-8205/727/1/L11}
}

@article{Osman2011Anisotropic,
  title = {Anisotropic Third-Moment Estimates of the Energy Cascade in Solar Wind Turbulence Using Multispacecraft Data},
  author = {Osman, K. T. and Wan, M. and Matthaeus, W. H. and Weygand, J. M. and Dasso, S.},
  journal = {Physical Review Letters},
  volume = {107},
  issue = {16},
  pages = {165001},
  numpages = {4},
  year = {2011},
  month = {Oct},
  publisher = {American Physical Society},
doi={10.1103/PhysRevLett.107.165001}
}

@article{Oughton2020CB,
year = {2020},
month = {jun},
publisher = {The American Astronomical Society},
volume = {897},
number = {1},
pages = {37},
author = {S. Oughton and W.H. Matthaeus},
title = {Critical Balance and the Physics of Magnetohydrodynamic Turbulence},
journal = {Astrophysical Journal},
doi={10.3847/1538-4357/ab8f2a}
}

@article{Oughton_1994, 
title={The influence of a mean magnetic field on three-dimensional magnetohydrodynamic turbulence}, 
volume={280}, 
journal={Journal of Fluid Mechanics}, 
author={Oughton, Sean and Priest, Eric R. and Matthaeus, William H.}, 
year={1994}, 
pages={95–117},
DOI      = {10.1017/S0022112094002867},
}

@BOOK{Parker-cmf,
    Bibkey   = {Parker-cmf},
    AUTHOR   = {Parker, E.N.},
    TITLE    = {Cosmical Magnetic Fields: {Their} Origin and Activity},
    YEAR     = {1979},
    Publisher = {Oxford University Press},
    Address  = {Oxford, UK},
    Comments = {don't have it}
}

@article{Percora2023PRL,
  title = {Three-Dimensional Energy Transfer in Space Plasma Turbulence from Multipoint Measurement},
  author = {Pecora, Francesco and Yang, Yan and Matthaeus, William H. and Chasapis, Alexandros and Klein, Kristopher G. and Stevens, Michael and Servidio, Sergio and Greco, Antonella and Gershman, Daniel J. and Giles, Barbara L. and Burch, James L.},
  journal = {Phys. Rev. Lett.},
  volume = {131},
  issue = {22},
  pages = {225201},
  numpages = {7},
  year = {2023},
  month = {Nov},
  publisher = {American Physical Society},
doi={10.1103/PhysRevLett.131.225201}
}

@article{Pecora2023ApJ,
author = {Pecora, Francesco and Servidio, Sergio and Primavera, Leonardo and Greco, Antonella and Yang, Yan and Matthaeus, William H.},
issn = {2041-8205},
journal = {The Astrophysical Journal Letters},
number = {2},
pages = {L20},
title = {{Multipoint Turbulence Analysis with HelioSwarm}},
volume = {945},
year = {2023},
doi={10.3847/2041-8213/acbb03}
}

@article{podesta2007anisotropic,
author = {Podesta, J.J.  and Forman, M.A.  and Smith, C.W. },
title = {Anisotropic form of third-order moments and relationship to the cascade rate in axisymmetric magnetohydrodynamic turbulence},
journal = {Physics of Plasmas},
volume = {14},
number = {9},
pages = {092305},
year = {2007},
doi={10.1063/1.2783224}
}

@article{politano1998karman,
  title={von K{\'a}rm{\'a}n--Howarth equation for magnetohydrodynamics and its consequences on third-order longitudinal structure and correlation functions},
  author={Politano, H. and Pouquet, A.},
  journal={Physical Review E},
  volume={57},
  number={1},
  pages={R21},
  year={1998},
  publisher={APS},
doi={10.1103/PhysRevE.57.R21}
}

@ARTICLE{   PolitanoPouquet98-grl,
    Bibkey   = {PolitanoPouquet98-grl},
    AUTHOR   = {Politano, H. and Pouquet, A.},
    TITLE    = {Dynamical length scales for turbulent magnetized flows},
    JOURNAL  = {Geophysical Research Letters},
    YEAR     = {1998},
    Volume   = {25},
    Pages    = {273--276},
    Comments = {an exact result for the scaling of a mixed third-order
                Elsasser structure fn; third moment},
doi={10.1029/97GL03642}
}

@Article{retino2022particle_PO, author={Retin{\`o}, Alessandro and Khotyaintsev, Yuri and Le Contel, Olivier and Marcucci, Maria Federica and Plaschke, Ferdinand and Vaivads, Andris and Angelopoulos, Vassilis and Blasi, Pasquale and Burch, Jim and De Keyser, Johan and Dunlop, Malcolm and Dai, Lei and Eastwood, Jonathan and Fu, Huishan and Haaland, Stein and Hoshino, Masahiro and Johlander, Andreas and Kepko, Larry and Kucharek, Harald and Lapenta, Gianni and Lavraud, Benoit and Malandraki, Olga and Matthaeus, William and McWilliams, Kathryn and Petrukovich, Anatoli and Pin{\c{c}}on, Jean-Louis and Saito, Yoshifumi and Sorriso-Valvo, Luca and Vainio, Rami and Wimmer-Schweingruber, Robert}, title={Particle energization in space plasmas: towards a multi-point, multi-scale plasma observatory}, journal={Experimental Astronomy}, year={2022}, month={12}, day={01}, volume={54}, number={2}, pages={427-471}, issn={1572-9508}, doi={10.1007/s10686-021-09797-7} }

@ARTICLE{   Robertson40,
    AUTHOR   = {Robertson, H. P.},
    TITLE    = {The Invariant Theory of Isotropic Turbulence},
    JOURNAL  = {Mathematical Proceedings of the Cambridge Philosophical Society},
    YEAR     = {1940},
    Volume   = {36},
    Pages    = {209--223},
    Comments = {isotropic tensors},
doi={10.1017/S0305004100017199}
}

@ARTICLE{   RobinsonRusbridge71,
    AUTHOR   = {Robinson, D. C. and Rusbridge, M. G.},
    TITLE    = {Structure of turbulence in the Zeta plasma},
    JOURNAL  = {Physics of Fluids},
    YEAR     = {1971},
    Volume   = {14},
    Pages    = {2499--2511},
    DOI      = {10.1063/1.1693359},
    adsurl   = {http://adsabs.harvard.edu/abs/1971PhFl...14.2499R},
    Comments = {}
}

@ARTICLE{   RuizEA11,
    AUTHOR   = {M. E. Ruiz and S. Dasso and W. H. Matthaeus and E. Marsch
                and J. M. Weygand},
    TITLE    = {Aging of anisotropy of solar wind magnetic fluctuations in
                the inner heliosphere},
    JOURNAL  = {Journal of Geophysical Research},
    YEAR     = {2011},
    Volume   = {116},
    DOI      = {10.1029/2011JA016697},
    EID      = {A10102},
    Comments = {perp and parallel lengthscale evolution; Helios},
}

@article{shebalin1983anisotropy,
  title={Anisotropy in MHD turbulence due to a mean magnetic field},
  author={Shebalin, J.V. and Matthaeus, W.H. and Montgomery, D.},
  journal={Journal of Plasma Physics},
  volume={29},
  number={3},
  pages={525--547},
  year={1983},
  publisher={Cambridge University Press},
doi={10.1017/S0022377800000933}
}

@article{Sorriso2007,
  title = {Observation of Inertial Energy Cascade in Interplanetary Space Plasma},
  author = {Sorriso-Valvo, L. and Marino, R. and Carbone, V. and Noullez, A. and Lepreti, F. and Veltri, P. and Bruno, R. and Bavassano, B. and Pietropaolo, E.},
  journal = {Phys. Rev. Lett.},
  volume = {99},
  issue = {11},
  pages = {115001},
  numpages = {4},
  year = {2007},
  month = {Sep},
  publisher = {American Physical Society},
doi={10.1103/PhysRevLett.99.115001}
}

@INPROCEEDINGS{SpenceEA19,
       author = {{Spence}, H.~E.},
        title = "{HelioSwarm: Unlocking the Multiscale Mysteries of Weakly-Collisional Magnetized Plasma Turbulence and Ion Heating}",
    booktitle = {AGU Fall Meeting Abstracts},
         year = 2019,
       volume = {2019},
        month = dec,
          eid = {SH11B-04},
        pages = {SH11B-04},
       adsurl = {https://ui.adsabs.harvard.edu/abs/2019AGUFMSH11B..04S}
}

@article{stawarz2009turbulent,
  title={The turbulent cascade and proton heating in the solar wind at 1 {AU}},
  author={Stawarz, J.E. and Smith, C.W. and Vasquez, B.J. and Forman, M.A. and MacBride, B.T.},
  journal={Astrophysical Journal},
  volume={697},
  number={2},
  pages={1119},
  year={2009},
  publisher={IOP Publishing},
doi={10.1088/0004-637X/697/2/1119}
}

@ARTICLE{TaylorEA03,
 title = {Recovering isotropic statistics in turbulence simulations: The Kolmogorov 4/5th law},
  author = {Taylor, Mark A. and Kurien, Susan and Eyink, Gregory L.},
  journal = {Physical Review E},
  volume = {68},
  issue = {2},
  pages = {026310},
  numpages = {8},
  year = {2003},
  month = {Aug},
  publisher = {American Physical Society},
  doi = {10.1103/PhysRevE.68.026310},
  url = {https://link.aps.org/doi/10.1103/PhysRevE.68.026310}
}

@article{tu1995mhd,
  title={{MHD} structures, waves and turbulence in the solar wind: Observations and theories},
  author={Tu, C and Marsch, E.},
  journal={Space Science Reviews},
  volume={73},
  number={1-2},
  pages={1--210},
  year={1995},
  publisher={Springer},
doi={10.1007/BF00748891}
}

@article{Verdini_2015,
year = {2015},
month = {may},
publisher = {The American Astronomical Society},
volume = {804},
number = {2},
pages = {119},
author = {A. Verdini and R. Grappin and P. Hellinger and S. Landi and W.C. Müller},
title = {Anisotropy of Third-Order Structure Functions in MHD Turbulence},
journal = {Astrophysical Journal},
doi={10.1088/0004-637X/804/2/119}
}

@article{wang2022strategies,
  title={Strategies for determining the cascade rate in MHD turbulence: isotropy, anisotropy, and spacecraft sampling},
  author={Wang, Y. and Chhiber, R. and Adhikari, S. and Yang, Y. and Bandyopadhyay, R. and Shay, M.A. and Oughton, S. and Matthaeus, W.H. and Cuesta, M.E.},
  journal={Astrophysical Journal},
  volume={937},
  number={2},
  pages={76},
  year={2022},
  publisher={IOP Publishing},
doi={10.3847/1538-4357/ac8f90}
}

@ARTICLE{   WeygandEA09,
    AUTHOR   = {J. M. Weygand and W. H. Matthaeus and S. Dasso and
                M. G. Kivelson and L. M. Kistler and C. Mouikis},
    TITLE    = {Anisotropy of the {Taylor} scale and the correlation
                scale in plasma sheet and solar wind magnetic field
                fluctuations},
    JOURNAL  = {Journal of Geophysical Research: Space Physics},
    YEAR     = {2009},
    Volume   = {114},
    DOI      = {10.1029/2008JA013766},
    EID      = {A07213},
    Comments = { }
}

@article{yang2021effects,
  title={Effects of forcing mechanisms on the multiscale properties of magnetohydrodynamics},
  author={Yang, Y. and Linkmann, M. and Biferale, L. and Wan, M.},
  journal={Astrophysical Journal},
  volume={909},
  number={2},
  pages={175},
  year={2021},
  publisher={IOP Publishing},
doi={10.3847/1538-4357/abdf58}
}

@ARTICLE{   YangEA24-effvisc,
    author = {Yang, Yan and Matthaeus, William H and Oughton, Sean and Bandyopadhyay, Riddhi and Pecora, Francesco and Parashar, Tulasi N and Roytershteyn, Vadim and Chasapis, Alexandros and Shay, Michael A},
    title = {Effective viscosity, resistivity, and Reynolds number in weakly collisional plasma turbulence},
    journal = {Monthly Notices of the Royal Astronomical Society},
    volume = {528},
    number = {4},
    pages = {6119-6128},
    year = {2024},
    month = {02},
    issn = {0035-8711},
    doi = {10.1093/mnras/stae355},
    url = {https://doi.org/10.1093/mnras/stae355},
    eprint = {https://academic.oup.com/mnras/article-pdf/528/4/6119/56707267/stae355.pdf},
}

@article{zikanov1998direct,
  title={Direct numerical simulation of forced MHD turbulence at low magnetic Reynolds number},
  author={Zikanov, O. and Thess, A.},
  journal={Journal of Fluid Mechanics},
  volume={358},
  pages={299--333},
  year={1998},
  publisher={Cambridge University Press},
doi={10.1017/S0022112097008239}
}

@ARTICLE{ZwebenEA79,
  title = {Small-Scale Magnetic Fluctuations Inside the Macrotor Tokamak},
  author = {Zweben, S. J. and Menyuk, C. R. and Taylor, R. J.},
  journal = {Phys. Rev. Lett.},
  volume = {42},
  issue = {19},
  pages = {1270--1274},
  numpages = {0},
  year = {1979},
  month = {May},
  publisher = {American Physical Society},
  doi = {10.1103/PhysRevLett.42.1270},
  url = {https://link.aps.org/doi/10.1103/PhysRevLett.42.1270}
}
\bibliographystyle{aasjournalv7}

\end{document}